\pdfoutput=1
\documentclass[aps,prd,superscriptaddress,notitlepage,11pt,final,longbibliography,nofootinbib]{revtex4-1}
\usepackage{graphicx}
\usepackage{amsfonts}
\usepackage{amsmath}
\usepackage{amsmath,amssymb,mathrsfs}
\usepackage[colorlinks=true, a4paper=true, pdfstartview=FitV, linkcolor=blue, citecolor=blue, urlcolor=blue]{hyperref}

\usepackage{subfigure}
\usepackage{epsfig}
\usepackage[usenames,dvipsnames]{xcolor}
\usepackage{bbm}
\usepackage{color}
\usepackage{amsmath}
\usepackage{natbib}
\usepackage{ulem}

\graphicspath{{./fig/}}

\newcommand{\be}{\begin{equation}}
\newcommand{\ee}{\end{equation}}
\newcommand{\bea}{\begin{eqnarray}}
\newcommand{\eea}{\end{eqnarray}}

\newcommand{\eqn}{\begin{eqnarray}}
\newcommand{\eqnx}{\end{eqnarray}}

\numberwithin{equation}{section}

\begin{document}

\title{Signum-Gordon shock waves in (2+1) and (3+1) dimensions}
\author{P. Klimas~}
\email{pawel.klimas@ufsc.br}
\affiliation{Departamento de F\'isica, Universidade Federal de Santa Catarina, Campus Trindade, 88040-900, Florian\'opolis-SC, Brazil}
\author{J. S. Streibel~}
\email{jstreibel@gmail.com}
\affiliation{Departamento de F\'isica, Universidade Federal de Santa Catarina, Campus Trindade, 88040-900, Florian\'opolis-SC, Brazil}

\begin{abstract}
This study introduces novel, exact solutions to the scalar field Signum-Gordon equation that feature a discontinuity near the light cone. These solutions, applicable in higher spatial dimensions ($n > 1$), extend previous limitations to one dimension. Our chosen ansatz leads to an ordinary equation with exact solutions obtained for $n = 2$ and $3$ spatial dimensions. The shock wave's energy trapped within the light cone is proportional to the wave's $n$-dimensional volume and the field discontinuity at the wavefront.
The investigation delves further into the behavior of shock waves when their driving force, represented by a delta function at the light cone, is disabled. Disabling this delta function disrupts energy transfer, preventing the wave's propagation as predicted by analytical calculations. We identify the region within the light cones where the field remains unaffected.
Two-dimensional ($n = 2$) simulations reveal the formation of intriguing structures upon source removal. These structures include a central, stable feature resembling an oscillon and a surrounding ring that breaks down into smaller oscillations. 
\end{abstract}


\maketitle

\section{Introduction}

One of the simplest scalar field models with non-analytic potential is the signum-Gordon (SG) model. The model's name originates from the fact that the force term in the equation of motion is determined by the sign function, $\frac{dV}{d\phi}=\lambda\; {\rm sgn}(\phi)$,  \cite{Arodz:2007ek}. 
The force in the SG model does not vanish for arbitrarily tiny variation from the minimum $\phi=0$, in contrast to the Klein-Gordon (KG) model, which has a quadratic potential, $\frac{dV}{d\phi}=m^2\phi$. This is why the force, which controls the release and propagation of tiny amplitude excitations from perturbed kinks and oscillons, is referred to as a {\it threshold force}. The SG potential $V(\phi)=\lambda\; |\phi|$ is piecewise linear. Despite its apparent simplicity, the model has surprisingly complex dynamics. For instance, the scattering of compact oscillons \cite{Hahne:2019ela} revealed the complex nature of the SG field dynamics.
In exploring the SG model, an essential aspect emerges in its relation to other models within the same category. This connection mirrors the link between the KG model and conventional field-theoretic models featuring parabolic minima. Given the quadratic nature of the KG potential, the solutions within this model serve as reliable approximations for small amplitude solutions in broader models exhibiting quadratic potential minima. Similarly, models characterized by non-analytic potentials at their minima share a comparable relationship with the SG model.  Small-amplitude oscillations dynamics are a universal phenomenon observable in various models within the same class, also referred to as models featuring V-shaped potentials, \cite{Arodz:2005gz}.

The example of this model type was presented in reference \cite{Arodz:2002yt}, focusing on a modified mechanical system involving inverted pendulums and its field theoretic limit.  Prior literature had already highlighted mechanical models with a limited number of degrees of freedom known to exhibit chaotic behavior arising from {\it grazing bifurcations}; references \cite{PhysRevA.27.1741, PhysRevE.49.1073, PhysRevE.50.4427} provide insight into this phenomenon. These models were also explored in Rodriguez-Coppola's 1992 study on the dynamics of electron gas in two dimensions \cite{Rodriguez_Coppola_1992} and in an article concerning plasma physics \cite{PhysRevLett.78.4761}. 

In the field theoretic limit detailed in reference  \cite{Arodz:2002yt}, the scenario involves an elastically interconnected chain of impacting pendulums within a gravitational field. These pendulums are constrained to swing within a specific angle domain $|\phi|<\phi_0$ due to the presence of rigid rods positioned at $\phi=\pm\phi_0$. 

\begin{figure}[h!]
\centering
\subfigure[]{\includegraphics[width=0.45\textwidth,height=0.2\textwidth, angle =0]{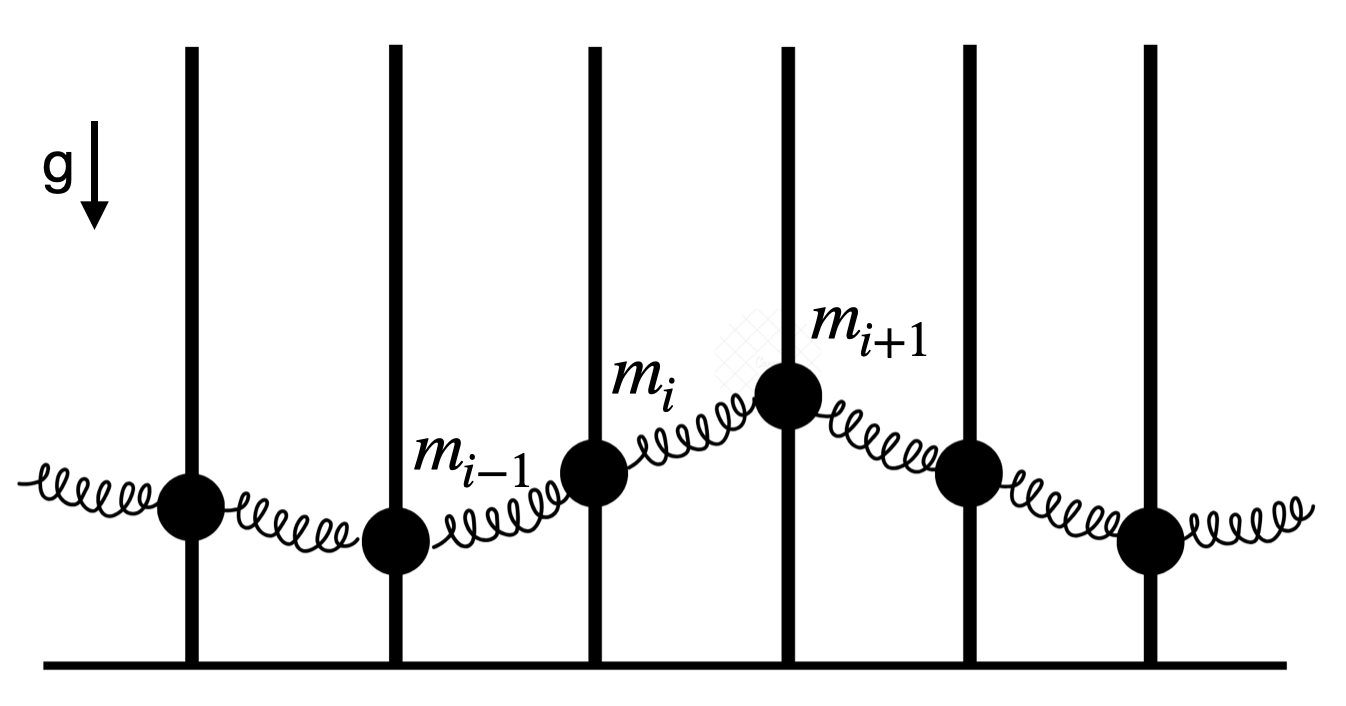}\label{fig:01a}}\hskip0.9cm
\subfigure[]{\includegraphics[width=0.45\textwidth,height=0.2\textwidth, angle =0]{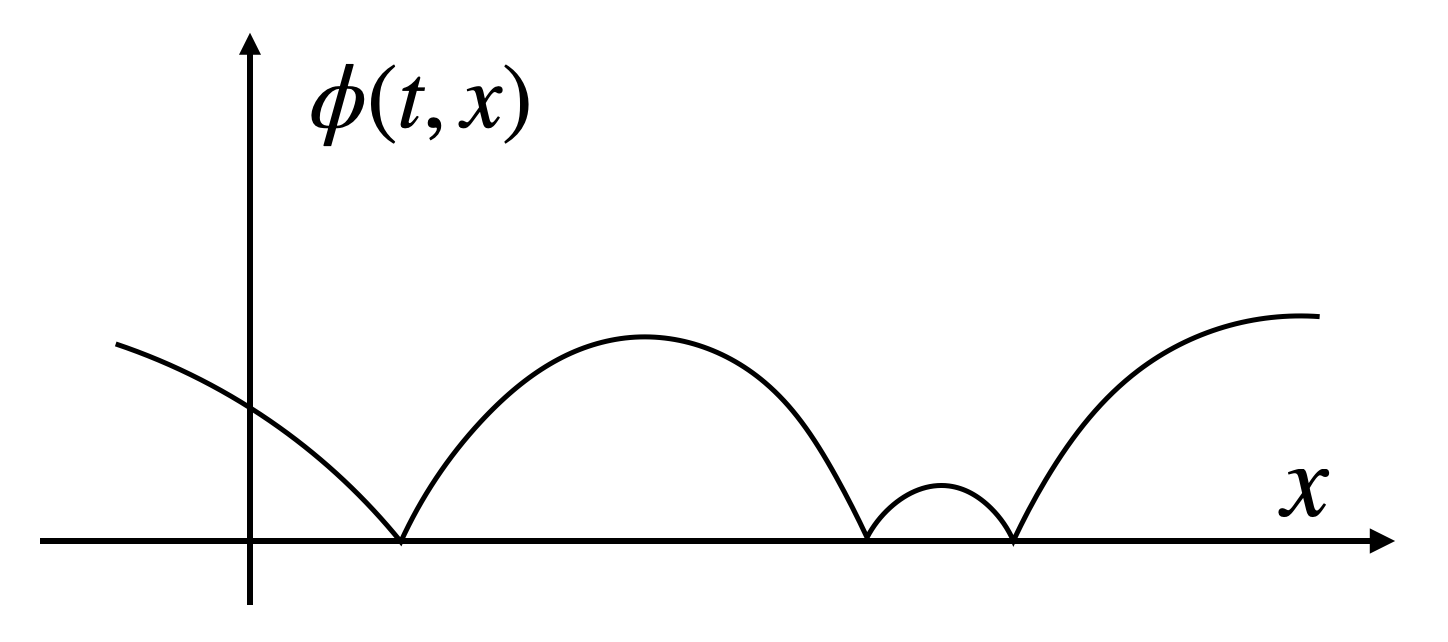}\label{fig:01b}}
\caption{The system comprises bouncing massive balls (a) interconnected by elastic springs (b), while the bouncing string, with vertical motion only, represents the continuous limit of the mechanical setup. Through a folding transformation, the auxiliary SG field is aligned with a solution derived from the mechanical string.}\label{fig:01}
\end{figure}

An alternative, simplified model is a discretized version analogous to the SG model. This model can be visualized as a series of massive balls bouncing along vertical rods, where each ball interacts with its neighbors through elastic springs, Fig.\ref{fig:01a}. 
The motion is limited to $\phi>0$ owing to a stiff floor at $\phi=0$, akin to a massive bouncing string in a continuous limit, as illustrated in Fig.\ref{fig:01b}.  Unlike the mechanical string, the SG field is not restricted by the stiff floor.  Interestingly, there exists a connection between the motion of a mechanical string and the SG field through a folding transformation, \cite{Arodz:2005gz}, eliminating a problematic reflection condition at $\phi=0$. This transformation correlates the evolution of the auxiliary SG field with the motion of the string.

\begin{figure}[h!]
\centering
{\includegraphics[width=0.45\textwidth,height=0.3\textwidth, angle =0]{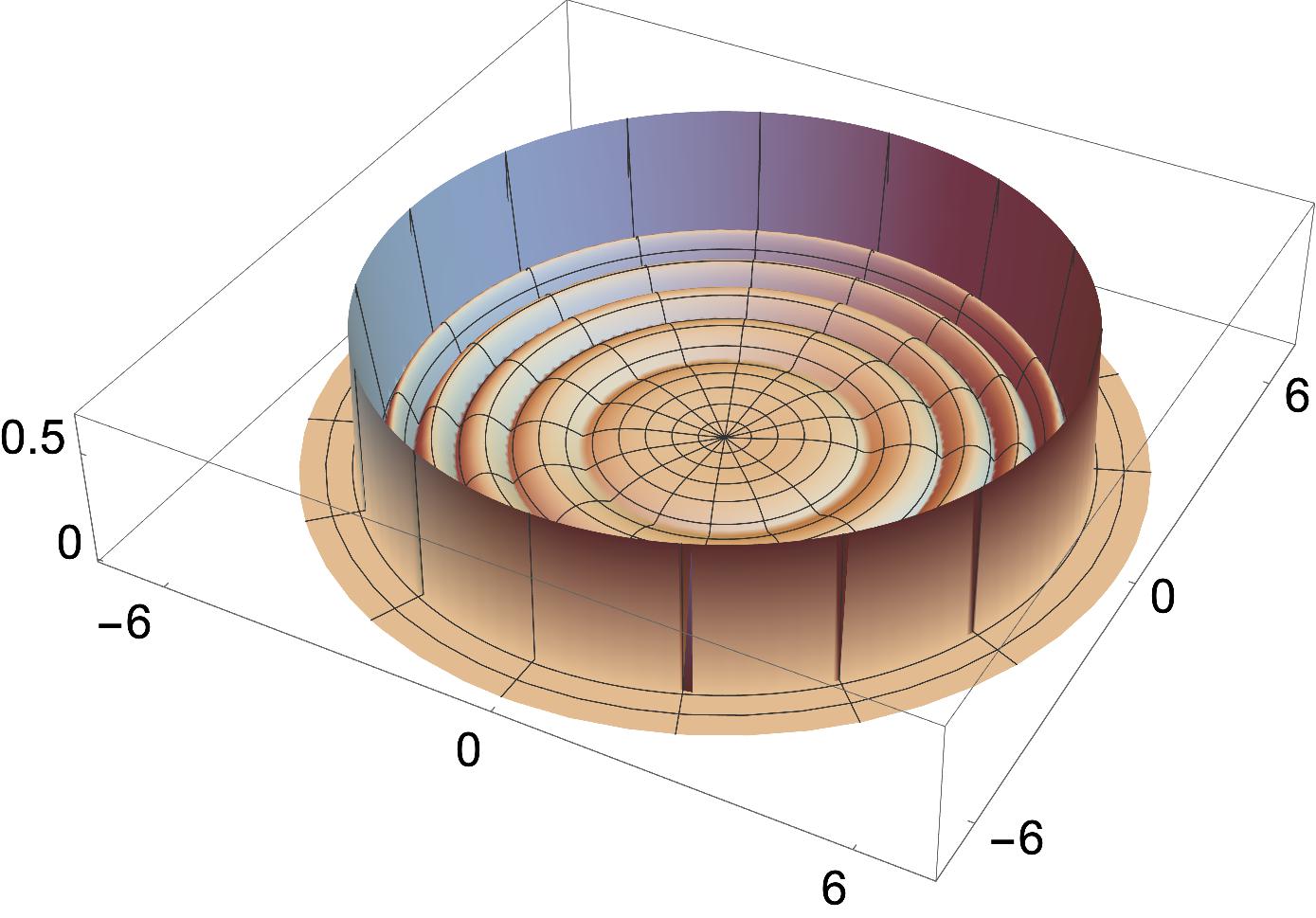}}
\caption{The bouncing membrane solution corresponds to the absolute value of the SG field $|\phi|$ as discussed in Section \ref{sec:sw2+1}.}\label{fig:02}
\end{figure}

To extend this model, the one-dimensional chain is replaced with a two-dimensional grid of balls, which, in a continuous representation, embodies a bouncing membrane. This grid can then be correlated with the SG field in 2+1 dimensions. The membrane solution, depicted in Figure \ref{fig:02}, illustrates shock waves investigated in Section \ref{sec:sw2+1}.

The question naturally arises whether models with V-shaped potentials can find application beyond the continuous limit of mechanical systems.  Intriguingly, the answer is affirmative. The presence of V-shaped potentials does not need to be assumed; these models can originate from more intricate physical theories that may not inherently include a potential. As stated in   \cite{Adam:2017srx}, the Skyrme model breaks down into two coupled BPS submodels, with one featuring a V-shaped potential. Consequently, solutions crafted for models with non-analytical potentials could potentially elucidate the behavior of the Skyrme field in instances of symmetry reduction.

The SG model exhibits several common features. One of these is the scaling symmetry, allowing for the existence of self-similar solutions \cite{Arodz:2007ek}. Another prominent characteristic of such models is the presence of {\it compactons}, referring to solutions with strictly compact support. This inclusion of compactons, or solutions with precisely compact support, is a distinguishing aspect of these models. Within these models, both topological compactons like the compact kink \cite{Arodz:2002yt} and non-topological compactons such as the exact compact oscillon \cite{Arodz:2007jh, Arodz:2011zm, ?wierczy?ski2021} or compact Q-balls \cite{Arodz:2008jk, Arodz:2008nm, Klimas:2017eft, Klimas:2018ywv} are feasible.
Moreover, a quadratic potential cannot effectively replicate the V-shaped potential in proximity to its minimum. This signifies that oscillations with moderate amplitudes inherently display nonlinearity. Essentially, in models featuring V-shaped potentials, {\it the conventional harmonic oscillator paradigm does not hold true}.

Our former study delves into {\it the shock wave solutions} within the SG model, focusing on reducing the model's partial differential equations to an ordinary equation to unveil a unique family of solutions. The analysis centers on the SG model in 1+1 dimensions, with the original exact solutions detailed in \cite{Arodz:2005gz}. These waves exhibit two fronts propagating in opposite directions at the speed of light, resulting in a nontrivial field within the light cone and a zero field outside of it, leading to discontinuity at the light cone's surface.
Furthermore, the SG model in 1+1 dimensions, incorporating a quadratic perturbation, yields exact shock wave solutions \cite{Klimas:2006fs}. Through recent numerical and analytical investigations, it was observed that non-exact configurations disintegrate into a sequence of compact oscillons \cite{Hahne:2019odw}. The study highlights that shock wave-like structures akin to these can emerge during the scattering process of compact oscillons, as evidenced in \cite{Hahne:2019ela}. This scattering mechanism serves as a conduit for transferring energy from specific SG configurations to smaller structures, giving rise to a radiation field. We posit that shock waves in higher spatial dimensions may play an analogous role, facilitating the transfer of energy from large-scale structures to small-scale structures. This constitutes a compelling motivation to further investigate such solutions.
 
This paper explores the extension of shock waves in the SG model to encompass two and three spatial dimensions through a blend of analytical and numerical methodologies. The investigation delves into solutions featuring discontinuities at the light cone. The structure of the paper is as follows: Section \eqref{exact_shockwaves} lays the groundwork by formulating a comprehensive analytic solution for the shock wave field and its total energy. Additionally, it delves into the delta force at the light cone, serving as a nonhomogeneous term in the SG equation. Subsequently, Section \eqref{sec:numeric} delves into the numerical resolution of a full-time dependent problem. Finally, Section \eqref{sec:comments} offers insights, comments, and reflections on the study.

\section{Exact shock waves} \label{exact_shockwaves}
\subsection{Shock wave ansatz in $(n + 1)$ dimensions}

In spaces with $(n + 1)$ dimensions, the  (SG) equation becomes:
\be
(\partial^2_t-\nabla^2)\phi(t,\vec x)+{\rm sgn}\,\phi(t,\vec x)=0.\label{eqsg1}
\ee
Here, the KG mass term is replaced by the signum function.
The vacuum configuration, characterized by $\phi=0$, represents a physically realizable field configuration (particularly evident in mechanical models).  To incorporate this case into the solution set of Eq. \eqref{eqsg1}, we adopt the convention that ${\rm sgn}(0)=0.$ We will continue using this definition throughout the study. 

Using spherical coordinates, Eq. \eqref{eqsg1} can be rewritten as:
\be
\left(\partial^2_t-\partial^2_r-\frac{n-1}{r}\partial_r-\frac{1}{r^2}\nabla^2_{S^{n-1}}\right)\phi(t,\vec x)+{\rm sgn}(\phi(t,\vec x))=0\label{eqsg2},
\ee
where  $r$  represents the radial coordinate $\sqrt{x_1^2+\ldots+x_n^2}$ and the operator $\nabla^2_{S^{n-1}}$  is determined by the set of angular variables. We seek solutions with spherical symmetry (independent of the angles).  Therefore,  $\nabla^2_{S^{n-1}}\phi(t,r)=0$, and the field equation simplifies to:
\be
\left(\partial^2_t-\partial^2_r-\frac{n-1}{r}\partial_r\right)\phi(t,r)+{\rm sgn}(\phi(t,r))=0.\label{phitr}
\ee
New variables, $x_{\pm}:=\frac{1}{2}(r\pm t)$, representing the coordinates of the light cone, can be introduced.  Consequently, the scalar field $\phi(x_+,x_-)$ satisfies the equation:
\be
-\partial_+\partial_-\phi-\frac{n-1}{2}\frac{\partial_+\phi+\partial_-\phi}{x_++x_-}+{\rm sgn}(\phi)=0.\label{phicone}
\ee
This equation has solutions that follow an ordinary differential equation. These solutions depend on a single variable, $z:=x_+x_-$, the product of light cone coordinates, and can be expressed as:
\be
\phi(x_+,x_-)=W(z)\qquad{\rm where}\qquad z=\frac{1}{4}(r^2-t^2).\label{ansatz}
\ee
Assuming this class of solutions, the partial differential equation \eqref{phicone}  simplifies to:
\be 
-zW''-\frac{n+1}{2}W'+{\rm sgn}(W)=0. \label{eq:generic-W}
\ee
We assume the solution is non-zero within the light cone $(z<0)$ and zero outside $(z>0)$. Since ${\rm sgn}(0)$ is defined as $0$, the constant solution $W(z)=0$  solves the equation for the region $z>0$. 

In this study, we investigate functions exhibiting discontinuities at the light cone $(r=t)$ due to our interest in shock waves. A reference to prior investigations of these shock wave solutions in one spatial dimension can be found in \cite{Arodz:2005bc}.  Following that approach, we examine expressions of the form: 
\be\label{eq:req-discont}
\phi(z)=\theta(-z)W(z)
\ee
where $\theta(-z)$ represents the Heaviside step function. We denote the left-hand side of the SG equation  \eqref{phicone} in variable $z$ as:
\be
F(\phi):=-\Big(z\phi'(z)\Big)'-\frac{n-1}{2}\phi'(z)+{\rm sgn}(\phi(z))\label{Fform}
\ee
Substituting \eqref{eq:req-discont} into \eqref{Fform} yields a Dirac delta term:\footnote{Consistent with \cite{vladimirov1971}, the action of a generalized function $f(z)$ on a test function $\varphi(z)$ is represented by $(f,\varphi)\equiv\int_{-\infty}^{\infty}dz f(z)\varphi(z)$. The Dirac delta distribution, in particular, acts on a test function according to $(\delta,\varphi)=\int_{-\infty}^{\infty}dz\;\delta(z)\varphi(z)=\varphi(0)$. }\footnote{This result can also be obtained using identities like 
\[
\theta'(-z)=-\delta(z),\quad z\delta(z)=0,\quad z\delta'(z)=-\delta(z),\quad {\rm sgn}(\theta(-z)W(z))=\theta(-z){\rm sgn}(W(z)).
\]
}
\be
(F(\phi),\varphi(z))=\left(C_n\,\delta(z),\varphi(z)\right),\qquad C_n\equiv \frac{n-1}{2}W(0).\label{eq:L}
\ee

Consistent with \cite{Arodz:2005bc}, expression \eqref{eq:req-discont} represents a solution to the homogeneous equation $F(\phi)=0$ when $n=1$, ($C_1=0$). This solution is derived by solving an ordinary differential equation \eqref{eq:generic-W} with the radial coordinate $r\ge 0$ replaced by a real number $x$. 

However, for $n\neq 1$, the term $F(\phi)$ does not vanish ($C_n\neq 0$).  Consequently, expression \eqref{eq:req-discont} is proportional to the {\it fundamental solution} $D(z)$ of the distributional equation:
\be
\Big(F(D(z)),\varphi(z)\Big)=(\delta(z),\varphi(z))\label{eqD}
\ee
where $\phi(z)=D(z)$ holds true for $W(0):=\frac{2}{n-1}$.

Similar to solutions observed in one spatial dimension ($n=1$) for the SG equation (or its perturbed forms \cite{Klimas:2006fs}), higher dimensional spaces ($n\ge 2$) can exhibit shock wave solutions with a fixed discontinuity value on the light cone surface, as demonstrated in \cite{Arodz:2005gz}. Analogous solutions might exist in higher dimensions if an additional force acts on the light cone surface. In other words, these solutions would satisfy a non-homogeneous SG equation Eq. \eqref{eq:L}  with a delta term proportional to the discontinuity value $W(0)$.

Furthermore, by allowing $W(0)$ to remain unfixed, we can identify a one-parameter family of solutions that generalize shock waves in (1+1) dimensions. The existence of such solutions in higher dimensions ($n\ge 2$) is a topic of current investigation. While numerical solutions are certainly attainable, a more insightful approach prioritizes the examination of analytical solutions first. We will provide a brief discussion on these novel solutions below, with further details available in sections \eqref{sec:sw2+1} and \eqref{sec:sw3+1}.

Shock waves are constructed from analytical solutions defined over small segments. By matching these solutions ("patches") for $z<0$, we can build the function $W(z)$. These constituent solutions are known as {\it partial solutions}.

To derive exact formulas for partial solutions, we assume $W(z)$ is a non-constant expression. In other words, it can only equal zero at specific, isolated points along the $z$-axis. Excluding these isolated points where $W(z)$ vanishes, we have either $W(z)$ greater than zero or $W(z)$ less than zero. In this context, it is convenient to rewrite the expression ${\rm sgn}(W)=\pm 1$ as ${\rm sgn}(W_k)=(-1)^k$, where $k=0,1,\ldots$ labels consecutive partial solutions $W_k$.  Under this notation, equation \eqref{eq:generic-W} transforms into a system of homogeneous linear equations:
\be
-zW_k''-\frac{n+1}{2}W_k'+(-1)^k=0.\label{eqWk}
\ee
This equation \eqref{eqWk} admits two classes of solutions:
\be \label{eq:sol-Wk-n=1}
W_k(z)=\alpha_k+\beta_k\ln(-z)+(-1)^k z \qquad {\rm for}\qquad n=1 
\ee
and
\be \label{eq:sol-Wk-n=23}
W_k(z)=\alpha_k+\beta_k(-z)^{-\frac{n-1}{2}}+\frac{2(-1)^k}{n+1}z, \qquad {\rm for}\qquad n\ge 2.
\ee
The first type of solution,  \eqref{eq:sol-Wk-n=1}, has been extensively studied in previous works \cite{Hahne:2019odw, Arodz:2005bc}. This paper focuses on the second type, Eq. \eqref{eq:sol-Wk-n=23}.

The free constant $\alpha_0=W_0(0)$ parametrizes the solutions. Setting $\beta_0=0$, we obtain a partial solution $W_0$ defined on the segment $-a_0<z<0$, where $a_0>0$.  For $n=1$, we find $\alpha_0=a_0$, and for $n\ge 2$, we have $\alpha_0=\frac{n+1}{2}a_0$. 

The solution $W_1(z)$ mates $W_0(z)$ at $z=-a_0$, meaning it satisfies both $W_1(-a_0)=0$ and $W'_1(-a_0)=W'_0(-a_0)$. These conditions allow us to calculate $\alpha_1$, $\beta_1$, and the next zero point $z=-a_1$, where $W_1(-a_1)=0$. This procedure can be iterated indefinitely. The complete solution is found by formulating and solving a set of recurrence relations. Sections \eqref{sec:sw2+1} and \eqref{sec:sw3+1} discuss a specific form of partial solutions, with illustrations provided in Figures \ref{fig:1} and \ref{fig:2}

\subsection{Shock wave energy} \label{sec:energy}
\subsubsection{Shock wave energy distribution}

As the shock wave propagates, the zone encompassing the excited scalar field expands with time. Notably, the wave's energy\footnote{Energy of post-wavefront oscillations} is confined entirely {\it within the light cone} ($r<t$). The integral form representing this energy is given by:
\be \label{eq:inner-cone}
E(t)=\int_{K(0,t)}d^nx\left[\frac{1}{2}(\partial_t \phi)^2+\frac{1}{2}(\partial_i \phi)^2+|\phi| \right].
\ee
Here, the solution's support $K(0,t)$ represents an $n$-sphere centered at $\vec r=0$ with radius $r$. It's important to note that this integral excludes the discontinuity of the wave, located at the light cone ($r=t$).

In summary, we calculate wave energy by segmenting it into three distinct zones delineated by the light cone:
\begin{enumerate}
\item $r<t$: This is the inner field region where energy exhibits well-behaved characteristics. Details are provided in Section \eqref{sec:inner-cone};
\item $r=t$: This zone encompasses the light cone surface, where the energy becomes undefined due to the field discontinuity and the leading delta term \eqref{eqD}. This is addressed in Section \eqref{sec:cone-surf};
\item $r>t$: This region represents the trivial vacuum outside the cone.
\end{enumerate}
While idealized exact solutions exhibit infinite gradient energy associated with the light cone discontinuity, this results in a significant discrepancy between numerical simulations and these idealized analytical solutions. Nonetheless, exact solutions with unbounded energy (like shock waves and self-similar solutions) hold relevance in physical applications. Certain physical processes, such as compact oscillon scattering \cite{Hahne:2019ela}, exhibit field configurations closely resembling such solutions, often over finite time intervals. In \cite{Hahne:2019odw}, we compared the radiation field produced during a scattering process to shock-like waves obtained for specific initial data. The remarkable similarity between the two configurations is noteworthy.

\subsubsection{Energy inside the light-cone} \label{sec:inner-cone}
Because in our ansatz $\phi$ depends only on time $t$ and radial distance $r$, expression \eqref{eq:inner-cone} reduces to:
\be 
E(t)=\int_{\mathcal{S}_n} d\Omega\int_0^t dr\,r^{n-1}\sum_{k=0}^{k_{\rm max}(t)}H_k(t,r)\label{energy-general}
\ee
where $W_k(z)$ partial solutions are provided by \eqref{eq:sol-Wk-n=1} and \eqref{eq:sol-Wk-n=23}. The contributions ($H_k(t, r)$) to energy density from partial solutions take the form:
\[
H_k(t,r)=\dfrac{r^2+t^2}{8}W_k'^2(z) + (-1)^kW_k(z), \qquad z=\frac{1}{4}(r^2-t^2),
\]
where $W'_k(z):=\frac{dW_k(z)}{dz}$.
In Eq. \eqref{energy-general}, $k_{\rm max}(t)$ represents the highest index of a partial solution $W_k(z)$ that can fit within the light cone at a given time $t$. The number of partial solutions constituting the shock wave at time $t$ is finite and increases with time. The integral  $\int_{\mathcal{S}_n} d\Omega$ denotes the area element of a unit sphere in $n$ spatial dimensions.

The expressions $W_k(z)$ and $W'_k(z)$ are non-zero only on the open segment $-a_k < z < -a_{k-1}$ but vanish outside this segment. Consequently, $W(-a_k)=0$. The algebraic representation of zeros $a_k$ varies with the number of spatial dimensions. For notational simplicity, we denote $i\equiv k_{\rm max}(t)$. The $i$-th zero appears at $r = 0$ at the instant $t_i=2\sqrt{a_i}$. In the interval $t_{i-1}<t<t_i$, the wave possesses $i + 1$ partial solutions identified by $k = 0, 1, ..., i.$

To explicitly show the contribution of each partial solution, the energy expression \eqref{energy-general} can be rewritten as:
\be
E(t)=\sum_{k=0}^iE_k(t) . \label{energy-nda}
\ee
Expressions $E_k(t)$, $k=0,1,\ldots,i-1$, represent the energies of the first $i$ partial solutions localized within shells with radii $c_{k}(t)\le r\le c_{k-1}(t)$, and $E_i(t)$ represents the energy of the central partial solution within the ball $0\le r\le c_{i-1}(t)$, where the scalar field's  radial running zeros, i.e. $\phi(t,r)|_{r=c_k(t)}=0$, take the form:
\be
c_k(t):=\sqrt{t^2-4a_k}. \label{ck}
\ee
The summation in Eq. \eqref{energy-nda} exhibits remarkable cancellation properties. As a consequence of these intricate cancellations, the expression ultimately simplifies to an elegant formula, as elaborated upon in subsequent sections. The final result is strikingly compact:
\[
E(t)=\int_{\mathcal{S}_n} d \Omega\int dr r^{n-1} [\phi]_{z=0}
\]
where $\int d \Omega$ represents the integral over the unit sphere in $n$ dimensions, and $[\phi]_z=0$ denotes the field discontinuity at the light cone ($r = t$). This remarkable outcome expresses the wave's energy within the light cone as the product of the $n$-dimensional volume (${\rm vol}_n(t)$) of a ball with radius $t$ and the field discontinuity at the light cone. Sections \eqref{sec:sw2+1} and \eqref{sec:sw3+1} explicitly demonstrate this reduction for $n = 2$ and $n = 3$, respectively.

Furthermore, it's noteworthy that the energy of shock waves in one spatial dimension can be formulated similarly. In this case, the expression reduces to $E(t) = 2t [\phi]_{z=0} = a_0 {\rm vol_1}(t)$, where  ${\rm vol_1}(t)=2t$ and $a_0$ is a constant value representing the field discontinuity. This observation highlights a key feature: the energy of a one-dimensional wave confined within the light cone increases proportionally to time $(E(t)\propto {\rm vol_1}(t))$. This proportionality between the wave's energy and the volume it occupies within the light cone appears to hold true for higher dimensions ($n = 2, 3$) as well, hence
\be \label{func:general_sw_energy}
E(t)={\rm vol}_n(t)\;[\phi]_{z=0},\qquad {\rm for} \qquad n=1,2,3.
\ee

\subsubsection{Energy  at the light cone's surface} \label{sec:cone-surf}

The preceding sections addressed the calculation of energy within the light cone ($r < t$). This section delves into the challenges of accurately estimating the energy precisely at the light cone's surface ($r = t$). This discussion is unavoidable for regularized models employing V-shaped potentials or their numerical implementations.

Shock waves exhibit an abrupt jump (discontinuity) in the scalar field ($\phi$) exactly at the light cone. This discontinuity possesses a significant challenge for calculating the energy density using standard methods. These methods typically rely on well-behaved, continuous functions for the field and its derivatives. 

Furthermore, the energy density for scalar field configurations involves the square of the field's gradient.  In the case of a shock wave, the gradient becomes proportional to a Dirac delta function ($\delta$-function) concentrated at the light cone surface.  However, the square of a distribution (like the Dirac delta) is not a mathematically well-defined operation. This lack of definition creates difficulties in directly calculating the energy density at the light cone. A strictly formal treatment yields an infinite value for the energy.

Although calculating the exact energy at the light cone surface is difficult, a technique called "regularization" can be employed to make the calculations more tractable. Regularization essentially modifies the delta term into a smoother function, enabling the definition of a well-behaved energy density. 

While regularization offers a way to address the discontinuity, it introduces additional considerations:
\begin{enumerate}
\item
{\it Non-convergence:} The regularization process introduces a parameter ($\varepsilon$) into the final energy expression. This parameter appears in the denominator, indicating that the energy calculation does not converge (reach a definitive value) as $\varepsilon$ approaches zero.
\item
{\it Smoothed discontinuity:} Regularization smooths out the sharp jump in the field ($\phi$) at the light cone, transforming it into a slope. This affects the gradient term, which contributes to the overall energy density.
\item
{\it Localized surface:} Due to regularization, the perfectly thin light cone surface becomes a narrow region with a thickness of $2\varepsilon$ centered around $r = t$. This introduces an additional dimension to the surface, making it $n$-dimensional instead of ($n-1$) dimensional (where $n$ is the number of spatial dimensions).
\item
{\it Finite reservoir and limited expansion:} Regularization introduces a finite, though exceptionally dense, energy reservoir at the light cone's surface. This reservoir acts as a source, supplying energy to the wave propagating within the light cone as it expands. In the absence of external forces (homogeneous equation), the wave can continue to expand as long as the reservoir possesses sufficient energy. This suggests that a shock wave formed from an initial discontinuity cannot indefinitely maintain its characteristics. Conversely, systems incorporating a delta function (or a regularized delta function) force at the light cone can exhibit regular expansion.
\item
{\it Wavefront size growth:}
One key distinction exists between wavefronts in one and higher spatial dimensions. In a single spatial dimension, the wavefront comprises two points. Conversely, in two dimensions, it forms a circle, and in three dimensions, a sphere. The size of this geometric feature (circle or sphere) increases with time as the wave expands. This implies that the energy density of a finite-energy initial discontinuity would become a less concentrated function as the wavefront expands (even neglecting the total energy transferred within the light cone). Notably, this effect of energy density spreading does not occur in one spatial dimension. This further underscores why regular oscillations are only observed in higher-dimensional systems with a delta force present.  The prior research \cite{Hahne:2019odw} investigated the initial stages of wave evolution in one spatial dimension ($n = 1$), specifically considering scenarios without an external delta function force at the light cone surface. 
\end{enumerate}

\subsection{Shock waves in ($2+1$) dimensions} \label{sec:sw2+1}
This section presents a detailed analysis of a novel, exact solution that characterizes shock waves. We will also examine the energy content of these waves, focusing specifically on the contribution from the region enclosed by the light cone.

\subsubsection{Partial solutions and recurrence relations}

The formula
\be \label{Wk-2d}
W_k(z)=\alpha_k+\frac{\beta_k}{\sqrt{-z}}+\frac{2}{3}(-1)^k z, \qquad{\rm where}\qquad z<0
\ee
provides the partial solutions for shock waves in two spatial dimensions. To ensure the absence of a singularity in the field at the light cone ($z = 0$), the parameter $\beta_0$ must be set to zero. The first zero ($z=-a_0$) of the solution is determined by the condition $W_0(-a_0)=0$, which implies  $\alpha_0=\frac{2}{3}a_0$, hence
\be \label{W0-2d}
W_0(z)=\frac{2}{3}(z+a_0).
\ee
The first zero  of the solution $W_0(z)$  determines the field discontinuity at the light cone ($z = 0$), denoted by
\be \label{phi2d-discontinuity}
[\phi]_{z=0}:=\lim_{z\rightarrow 0-}\phi(z)-\lim_{z\rightarrow 0+}\phi(z)=\lim_{z\rightarrow 0-}W_0(z)=\frac{2}{3}a_0.
\ee
Alternatively, the field discontinuity at the light cone can be treated as an independent parameter that dictates the value of $a_0$. By employing recurrence relations, all other parameters and zeros of the solution can then be determined. Notably, the initial zero ($z=-a_0$) coincides with a zero of $W_1(z)$. This allows for the elimination of $\alpha_1$ by expressing it in terms of $a_0$. Similarly, the constraint $W_k(-a_{k-1})=0$ facilitates the elimination of subsequent $\alpha_k$ parameters. It provides
\be
W_k(z)=(-1)^k\left[\frac{2}{3}(z+a_{k-1})+b_k\left(\frac{1}{\sqrt{-z}}-\frac{1}{\sqrt{a_{k-1}}}\right)\right], \qquad k=1,2,\ldots \label{Wk2d}
\ee
where $b_k\equiv (-1)^k\beta_k$. 

\begin{figure}[h!]
\centering
\subfigure[]{\includegraphics[width=0.45\textwidth,height=0.3\textwidth, angle =0]{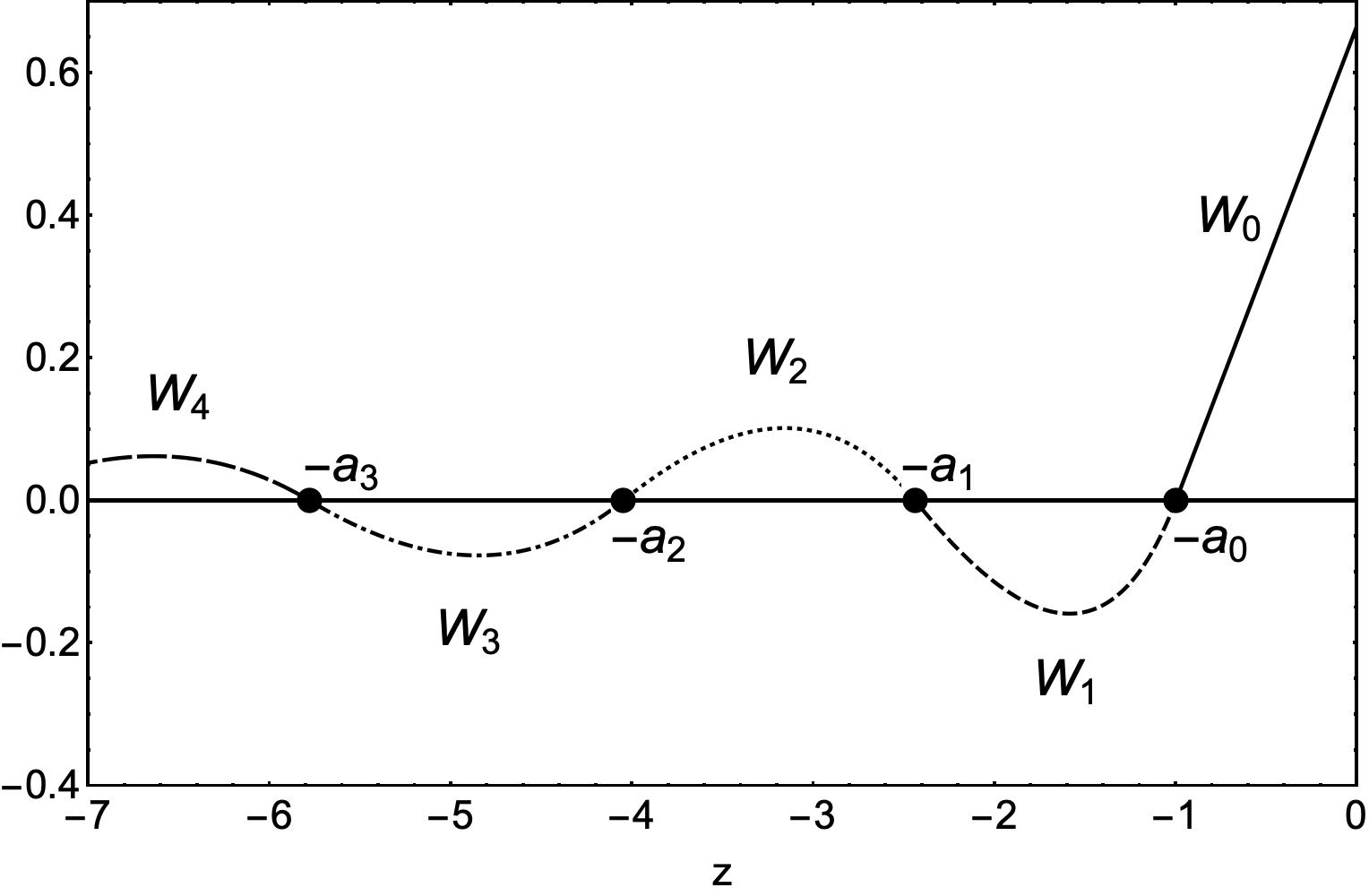}\label{fig:1a}}\hskip0.9cm
\subfigure[]{\includegraphics[width=0.35\textwidth,height=0.3\textwidth, angle =0]{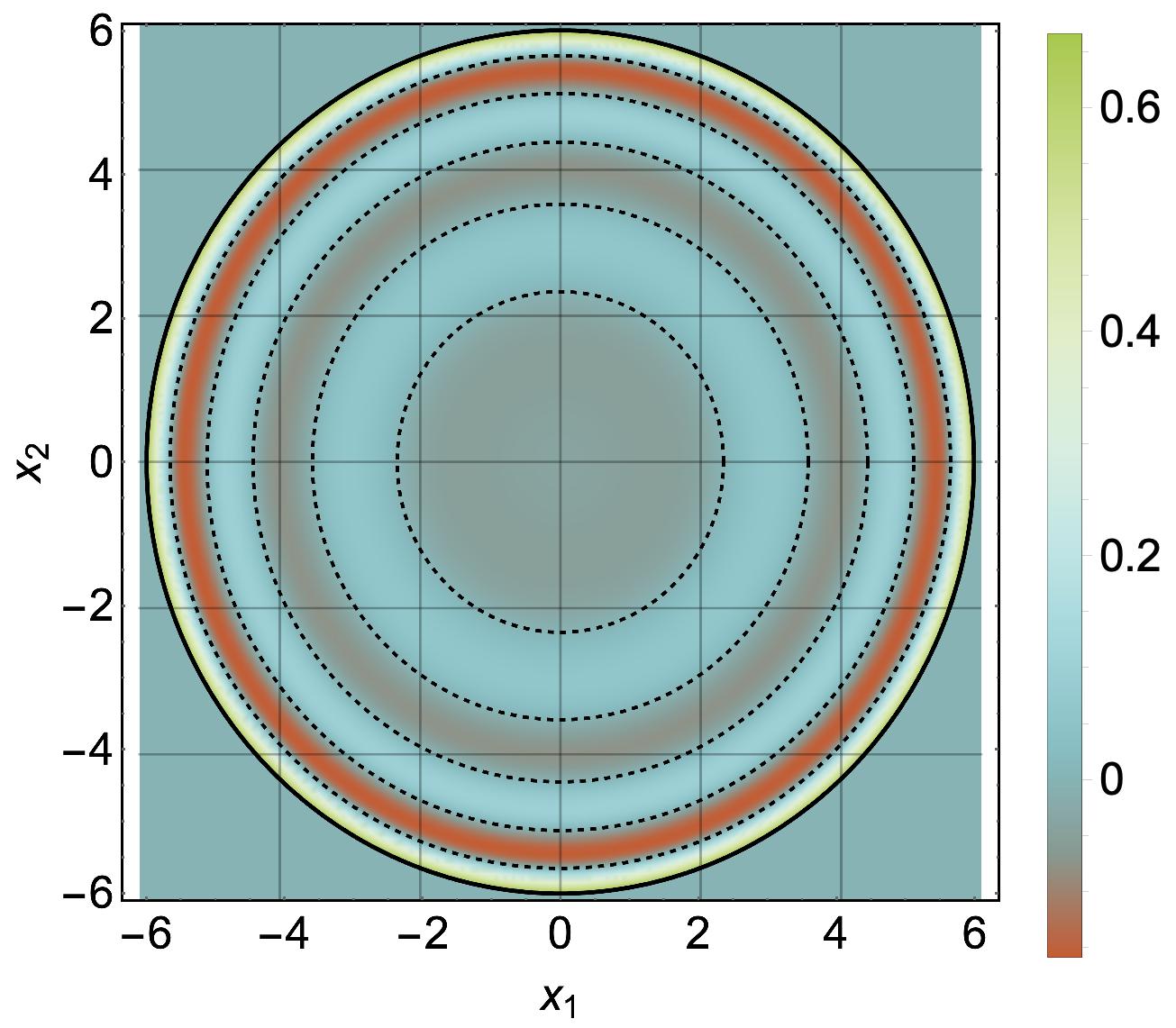}\label{fig:1b}}
\caption{(a) Partial solutions $W_k(z)$  and (b) the shock wave at $t=6.0$ in two spatial dimensions.}\label{fig:1}
\end{figure}

The first derivative, $W'(z)$, of the solution $W(z)$ exhibits continuity at corresponding points. Mathematically, this translates to $W'_k(-a_{k-1})=W'_{k-1}(-a_{k-1})$, where the prime notation denotes differentiation. This continuity requirement leads to a set of recurrence relations for the $b_k$ coefficients. These relations are expressed as 
\be \label{bk-recursion-relation}
b_k=-b_{k-1}-\frac{8}{3}a_{k-1}^{3/2},
\ee
with the initial condition $b_0 = 0$. Consequently, the $b_k$ coefficients can be determined through a summation involving all preceding parameters ($a_n$) where $n$ is less than $k$:
\be 
b_k=\frac{8}{3}\sum_{n=0}^{k-1}(-1)^{n+k}a_n^{3/2},\qquad k=1,2,\ldots.
\ee
The $k-$th zero of the solution, denoted by $z=-a_k$, is obtained by solving the equation $W_k(-a_k)=0$. This equation can be recast as a second-order equation
\be
(\sqrt{a_k})^2+\sqrt{a_{k-1}}\sqrt{a_k}+\frac{3}{2}\frac{b_k}{\sqrt{a_{k-1}}}=0\qquad k=1,2,\ldots\label{akeq}
\ee
for the square root of $\sqrt{a_k}$.
Solving the equation \eqref{akeq} for the square root of $a_k$ and squaring the resulting expression leads to
\be
a_k=s_k a_{k-1}=\left(\prod_{i=0}^{k}s_i\right)a_0,\qquad k=1,2,\ldots
\label{eq:linear-ak}
\ee
where
\begin{align}
&s_0\equiv 1,\\
&s_k\equiv\frac{1}{2}\Big(1-\xi_k-\sqrt{1-2\xi_k}\Big),\qquad k=1,2,\ldots.\label{ak2d}
\end{align}
The expression $\xi_k$ in \eqref{ak2d} is as follows:
\be
\xi_k\equiv8\sum_{n=0}^{k-1}(-1)^{n+k}\left(\frac{a_n}{a_{k-1}}\right)^{3/2}=8\sum_{n=0}^{k-1}(-1)^{n+k}\left(\frac{\prod_{i=0}^ns_i}{\prod_{j=0}^{k-1}s_j}\right)^{\frac{3}{2}}.\label{xik}
\ee
It is worth noting that each zero $a_k$ is proportionate to $a_0$.
We can express the partial solutions, denoted by $W_k(z)$, in a specific form 
\begin{align}
W_0(z)&=\frac{2}{3}(z+a_0),\\
W_k(z)&=\frac{2}{3}(-1)^{k}\frac{\sqrt{d_k}+\sqrt{-z}}{\sqrt{-z}}\Big(\sqrt{a_k}-\sqrt{-z}\Big)\Big(\sqrt{-z}-\sqrt{a_{k-1}}\Big),\qquad k=1,2.\ldots
\end{align}
This form clarifies the zeros of these solutions. We achieve this by defining an expression, denoted by
\be
d_k\equiv \frac{1}{2}\Big(1-\xi_k+\sqrt{1-2\xi_k}\Big)a_{k-1},\qquad k=1,2,\ldots\nonumber
\ee
where a specific condition holds $\sqrt{a_{k-1}}<\sqrt{-z}<\sqrt{a_k}$. The recurrence relations described in equations \eqref{eq:linear-ak}-\eqref{xik} determine the zeros of the function $W(z)$. This solution demonstrates the possibility of analytically determining zeros in higher dimensions, unlike the one-dimensional case where numerical methods are typically required.  Figure \ref{fig:1} depicts both the function $W(z)$ and the shock wave in a two-dimensional space.
\subsubsection{Shock wave energy in (2+1) dimensions} \label{sec:en2+1}

In two spatial dimensions, equations \eqref{W0-2d} and \eqref{Wk-2d} provide the expressions for $W_k(t)$. The $W'_k(z)$ expressions read
\begin{align}
	W'_0(z)&=\dfrac23 \\
	W'_k(z)&=(-1)^k\left(\dfrac23+\dfrac{b_k}{2(-z)^{3/2}}\right), \qquad k=1,2,\ldots.
\end{align}
The energy \eqref{energy-general} of a solution  is given by 
\be
E(t)=2\pi\Big[\int_0^{c_{i-1}}dr\; r\, H_i(t,r)+\sum_{k=1}^{i-1}\int_{c_k}^{c_{k-1}}dr \;r\,H_k(t,r)+\int_{c_0}^t dr \;r\,H_0(t,r)\Big]\label{thigh}
\ee
where $t>2\sqrt{a_{i-1}}$. The summation represented by equation \eqref{thigh} contains only one term 
\be
E(t)=2\pi \int_{0}^t dr \;r\,H_0(t,r)=\pi t^2 \frac{2a_0}{3}= {\rm vol}_{2}(t) [\phi]_{z=0}\label{tsmall}
\ee
when $t<2\sqrt{a_0}$. This condition holds true before the occurrence of $c_0(t)$.

We now aim to demonstrate that equation \eqref{thigh} leads to an expression directly comparable to $ {\rm vol}_{2}(t) [\phi]_{z=0}$. Consequently, we can rewrite the integrals in equation \eqref{thigh} as follows
\begin{align}
2\pi \int_{0}^{c_{i-1}}dr\; r\, H_i(t,r)&\equiv X_i\nonumber \\
2\pi \int_{c_k}^{c_{k-1}}dr\; r\, H_k(t,r)&\equiv\overbrace{2\pi \int_{c_k}^{0}dr\; r\, H_k(t,r)}^{Y_k(t)}+\overbrace{2\pi \int_0^{c_{k-1}}dr\; r\, H_k(t,r)}^{X_k(t)}\nonumber \\
2\pi \int_{c_0}^{t}dr\; r\, H_0(t,r)&\equiv\overbrace{2\pi \int_{c_0}^{0}dr\; r\, H_0(t,r)}^{Y_0(t)}+\overbrace{2\pi \int_0^{t}dr\; r\, H_0(t,r)}^{X_0(t)}.\nonumber
\end{align}
The sum of the terms can be expressed in the following form
\be
E(t)=X_i(t)+\sum_{k=0}^{i-1}\Big(X_k(t)+Y_k(t)\Big)=X_0(t)+\sum_{k=1}^i\Big(X_k(t)+Y_{k-1}(t)\Big).\label{energysum}
\ee
Once we substitute the analytical solution into these formulas, we arrive at
\begin{align}
X_0(t)&=\frac23a_0\pi t^2,\label{X0} \\
X_k(t) &= \frac{\pi}{72}\frac{c_{k-1}^2}{a_{k-1}^2}\left[3b_k+4a_{k-1}^{3/2}\right]^2\label{Xk},\qquad k=1,2,\ldots \\
Y_k(t) &= -\frac{\pi}{72}\frac{c_{k}^2}{a_{k}^2}\left[48a_{k-1}a_k^2-32a_k^3+96b_ka_k^{3/2}-72a_k^2\frac{b_k}{\sqrt{a_{k-1}}}+9b_k^2\right] \label{Yk}, \qquad k=0,1,2,\ldots
\end{align}
Equation \eqref{Yk} can be expressed in an alternative form by leveraging the relationships established in equations \eqref{bk-recursion-relation} and $\frac{b_k}{\sqrt{a_{k-1}}}=-\frac{2}{3}[a_k+\sqrt{a_{k-1}}\sqrt{a_k}]$ which follows from \eqref{akeq}. After some algebraic steps, we arrive at
\begin{align}
Y_k(t)=&-\frac{\pi}{72}\frac{c_{k}^2}{a_{k}^2}\Big[3b_{k+1}+4a_k^{3/2}\Big]^2\label{Yka}\\
&+\frac{3\pi}{2}c_k^2\Big[a_{k+1}+a_{k-1}-4a_k+\sqrt{a_k}\Big(\sqrt{a_{k+1}}+\sqrt{a_{k-1}}\Big)\Big].\label{Ykb}
\end{align}
The first term, denoted by equation \eqref{Yka}, is equal to $-X_{k+1}$. In contrast, the second term, given by equation, \eqref{Ykb}, is proportional to $W_k(-a_k)$. Since $z=-a_k$ is a zero of the function $W_k(z)$,  this second term becomes zero. Indeed, substituting $z=-a_k$ into equation \eqref{Wk2d} results in
\[
-a_k+a_{k-1}+\frac{3}{2}\frac{b_k}{\sqrt{a_k}}-\frac{3}{2}\frac{b_k}{\sqrt{a_{k-1}}}=0.
\]
In the following step, we substitute $b_{k+1}$ for $b_k$ using the relation established in equation \eqref{bk-recursion-relation}. This substitution leads to the elimination of the terms $\frac{b_{k+1}}{\sqrt{a_k}}$ and $\frac{b_{k}}{\sqrt{a_{k-1}}}$ due to the relationship given in equation  \eqref{akeq}. The result of these substitutions is
\[
a_{k+1}+a_{k-1}-4a_k+\sqrt{a_k}\Big(\sqrt{a_{k+1}}+\sqrt{a_{k-1}}\Big)=0.
\]
This result establishes the following identity
\be
X_{k+1}(t)+Y_{k}(t)=0, \qquad k=0,1,2,\ldots
\ee
Finally, equation \eqref{energysum} epresses the energy, $E(t)$, of the (2+1)-dimensional shock wave:
\begin{align}
E(t)= \frac23\pi a_0\,t^2 = {\rm vol}_2(t)\,[\phi]_{z=0}\label{energy2dim}
\end{align}
where ${\rm vol}_2(t)=\pi\,t^2$ and the discontinuity term $[\phi]_{z=0}$ is given by the  expression \eqref{phi2d-discontinuity}.

\subsection{Shock waves in ($3+1$) dimensions} \label{sec:sw3+1}
\subsubsection{Partial solutions and recurrence relations}
The partial solutions for shock waves in three dimensions ($n = 3$) read
\be
W_k(z)=\frac{1}{2}(-1)^kz+\alpha_k+\frac{\beta_k}{z}
\ee
where the parameter $\beta_0$ equals zero. The condition $W_k(-a_{k-1})=0$ allows us to eliminate $\alpha_k$, resulting in
\begin{align}
W_0(z)&=\frac{1}{2}(z+a_0),\label{w0}\\
W_{k}(z)&=(-1)^k\left[\frac{1}{2}(z+a_{k-1})+b_{k}\left(\frac{1}{z}+\frac{1}{a_{k-1}}\right)\right],\qquad k=1,2,\ldots\label{wk}
\end{align}
where $\beta_k=(-1)^kb_k$. The  partial solutions $W_{k-1}(z)$ and $W_{k}(z)$, fulfill a  matching condition at $z = -a_{k-1}$, expressed as $W'_{k}(-a_{k-1})=W'_{k-1}(-a_{k-1})$. This condition leads to a recurrence relation
\be
b_k=-b_{k-1}+ a^2_{k-1}, \nonumber
\ee
where $ b_0=0$, which has solution
\be
b_k=-\sum_{n=0}^{k-1}(-1)^{n+k} a^2_{n}. \label{betak3d}
\ee
Finally, we enforce the condition $W_k(-a_k)=0$. This leads to the recurrence relation  $a_k=\frac{2 b_k}{a_{k-1}}$. Combining this with the previously derived relation \eqref{betak3d} for $b_k$, we arrive at a new expression for $a_k$
\be
a_k=\frac{2}{a_{k-1}}\sum_{n=0}^{k-1}(-1)^{n+k-1} a^2_{n}.\label{recurrenceak}
\ee
The recurrence relation presented in equation \eqref{recurrenceak} has the following simple solution
\be \label{eq:a_k-exact}
a_k=(k+1)a_0.
\ee 
The solution can be verified by substituting it back into equation \eqref{recurrenceak} and utilizing the  formula  $$\sum_{m=1}^k(-1)^{m+k}m^2=\frac{k(k+1)}{2}.$$ 
\begin{figure}[h]
\centering
\subfigure[]{\includegraphics[width=0.45\textwidth,height=0.3\textwidth, angle =0]{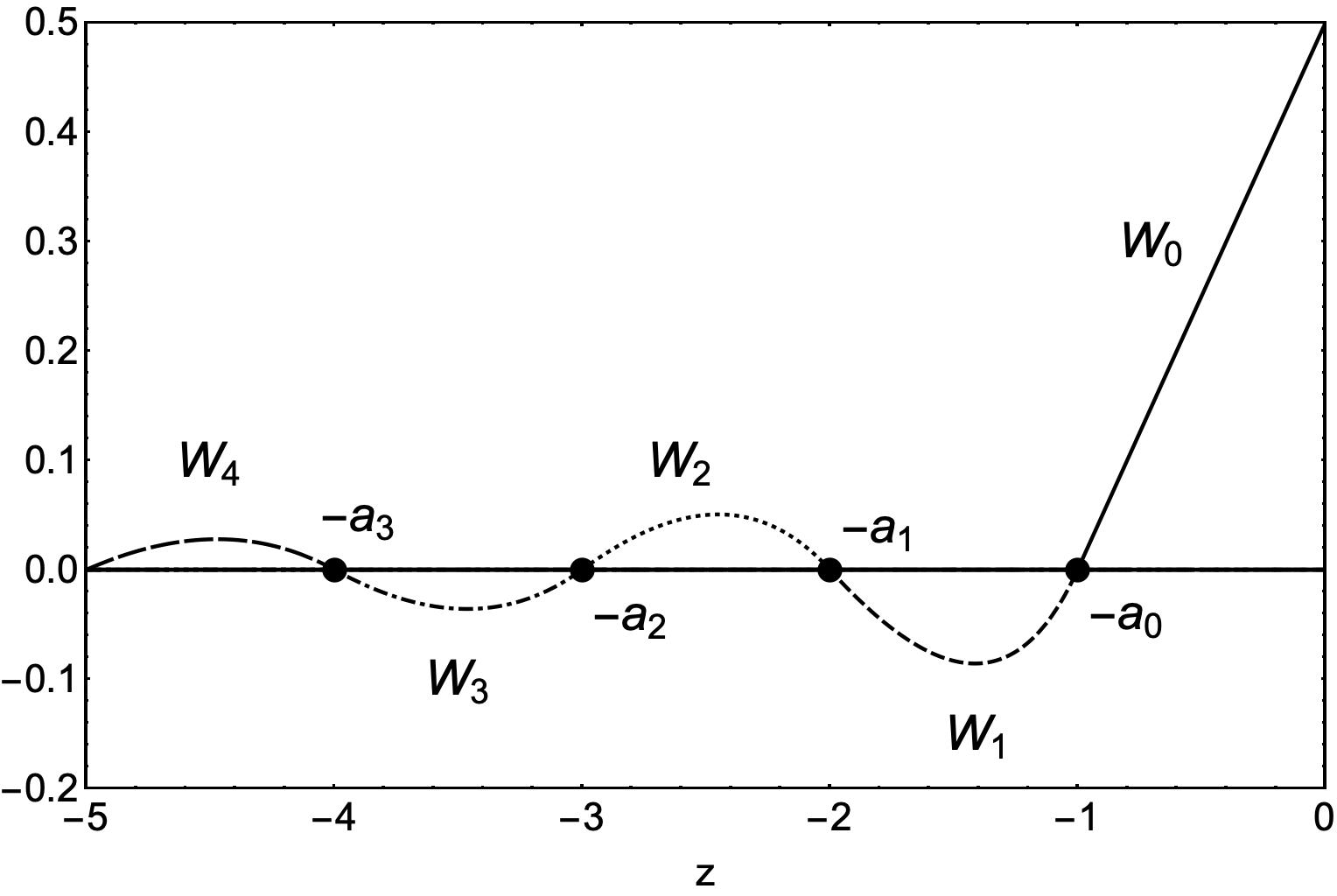}\label{fig:2a}}\hskip0.9cm
\subfigure[]{\includegraphics[width=0.35\textwidth,height=0.3\textwidth, angle =0]{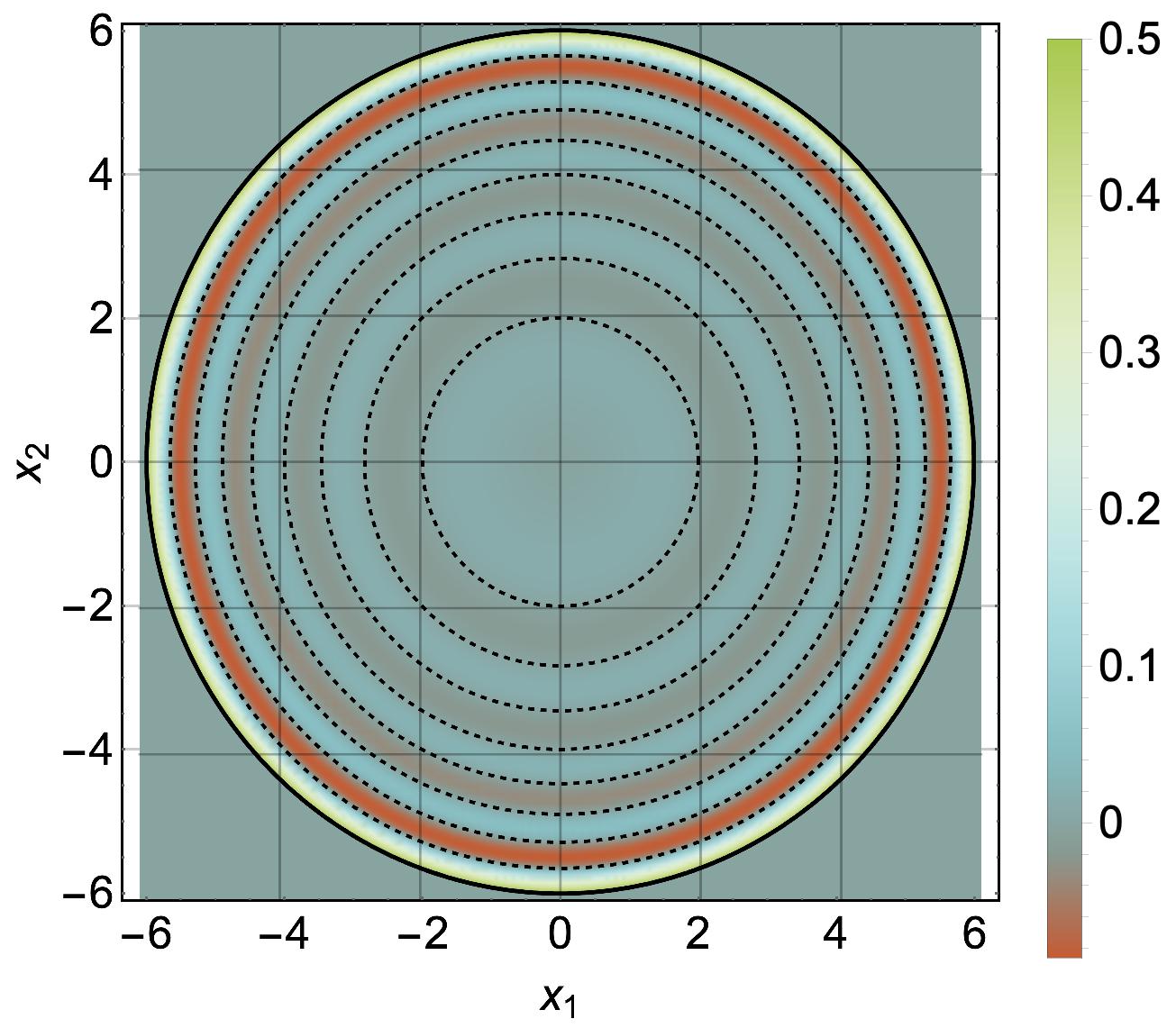}\label{fig:2b}}
\caption{(a) Partial solutions $W_k(z)$ and (b) the shock wave section $x^3=0$ at $t=6.0$ in three spatial dimensions.}\label{fig:2}
\end{figure}

The ability to determine the exact locations of the zeros proves to be very valuable.  This allows us to derive explicit expressions for the partial solutions
\be
W_k(z)=\frac{(-1)^k}{2z}\Big(z+ka_0\Big)\Big(z+(k+1)a_0\Big),\qquad k=0,1,2,\ldots.\label{Wk3d}
\ee 
Figure \ref{fig:2a} illustrates a selection of the initial partial solutions, denoted by $W_k(z)$.

\subsubsection{Shock wave energy in (3+1) dimensions} \label{sec:en3+1}
The energy of a shock wave in (3+1) dimensions is expressed by the formula
\eqref{energysum},  where
\begin{align}
X_0(t)&\equiv 4\pi \int_0^{t}dr\; r^2\, H_0(t,r),\label{x0}\\
X_k(t)&\equiv 4\pi \int_0^{c_{k-1}}dr\; r^2\, H_k(t,r),\qquad k=1,2,\ldots\label{xk}\\
Y_k(t)&\equiv 4\pi \int_{c_k}^{0}dr\; r^2\, H_k(t,r),\qquad k=0,1,2,\ldots.\label{yk}
\end{align}

Substituting the expression for $W_k$ provided in equation  \eqref{Wk3d} and the specific form for $W'_k(z)$,
\be
W'_k(z)=\frac{(-1)^k}{2z^2}\Big(z^2-k(k+1)a_0^2\Big),\label{dWk3d}
\ee 
 into formulas \eqref{x0},  \eqref{xk} and \eqref{yk} yields
\begin{align}
X_0(t)&=4\pi\left(\frac{a_0}{6}t^3\right),&&\label{form1}\\
X_k(t)&=4\pi\left(\frac{a_0}{24k}c_{k-1}^3(t)\right),&&k=1,2,\ldots\label{form2}\\
Y_k(t)&=4\pi\left(-\frac{a_0}{24(k+1)}c_k^3(t)\right), &&k=0,1,2,\ldots.\label{form3}
\end{align}
Based on the relationships established in equations \eqref{form2} and \eqref{form3}, we can determine that $X_k(t)=-Y_{k-1}(t)$.  This consequence simplifies the energy formula \eqref{energysum}  in $(3+1)$ dimensions to $E(t)=X_0(t)$, which holds true for any value of $t$.
This result implies that the energy of the wave within the light cone increases as a cubic function of time. The discontinuity of the field at the light cone in three spatial dimensions is described by
\be
[\phi]_{z=0}=\lim_{z\rightarrow 0-}W_0(z)=\frac{a_0}{2}.
\ee
By considering the volume of the sphere occupied by the wave, denoted as ${\rm vol}_3(t)$ and defined as $\frac{4\pi}{3}t^3$, we can relate the wave's energy to this volume and the strength of the discontinuity. This relationship expresses the wave's energy as
\be
E(t)={\rm vol}_3(t)[\phi]_{z=0}.
\ee
In conjunction with equation \eqref{energy2dim}, this result completes the proof of formula \eqref{func:general_sw_energy}.

\subsection{Numerical approximation of the solution with Dirac delta force}\label{subs:delta-force}

In Section  \eqref{exact_shockwaves}, we established that a solution exists for the ordinary differential equation involving a Dirac delta force. This solution describes the propagation of a field configuration exhibiting a persistent discontinuity at the light cone. The equation has the form $F(\phi(z))=C_n\delta(z)$, where $F(\phi)$ is defined by equation \eqref{Fform} and $C_n\equiv\frac{n-1}{2}W(0)$. The general form of solution is presented in equation \eqref{eq:req-discont}, which utilizes the Heaviside step function.

This section aims to obtain a numerical approximation for the solution $\phi(z)$ using numerical methods in response to the Dirac delta force $\delta(z)$.

Before delving into the numerical solution of our problem, let's revisit a well-known case: the harmonic oscillator. This example exhibits some key similarities to the problem we will address. However, due to its relative simplicity, it serves as a well-suited introduction to the subject matter. The harmonic oscillator fundamental solutions satisfy the equation $D''(t)+\omega^2D(t)=\delta(t)$. These solutions fall into two categories: retarded (denoted by $D_R(t)$) and advanced (denoted by $D_A(t)$).
To obtain numerical solutions, we need to replace the Dirac delta function with an approximation that can be handled computationally. A common approach is to use a regularized version, such as the one shown in equation
\be
\delta_{\epsilon}(t)=\frac{1}{\sqrt{2\pi\epsilon}}e^{-\frac{t^2}{2\epsilon}}.\label{regularizeddelta}
\ee
This approximation approaches the true Dirac delta function as the parameter $\varepsilon$ tends to zero.

\begin{figure}[h!]
\subfigure[]{\includegraphics[width=0.45\textwidth,height=0.25\textwidth, angle =0]{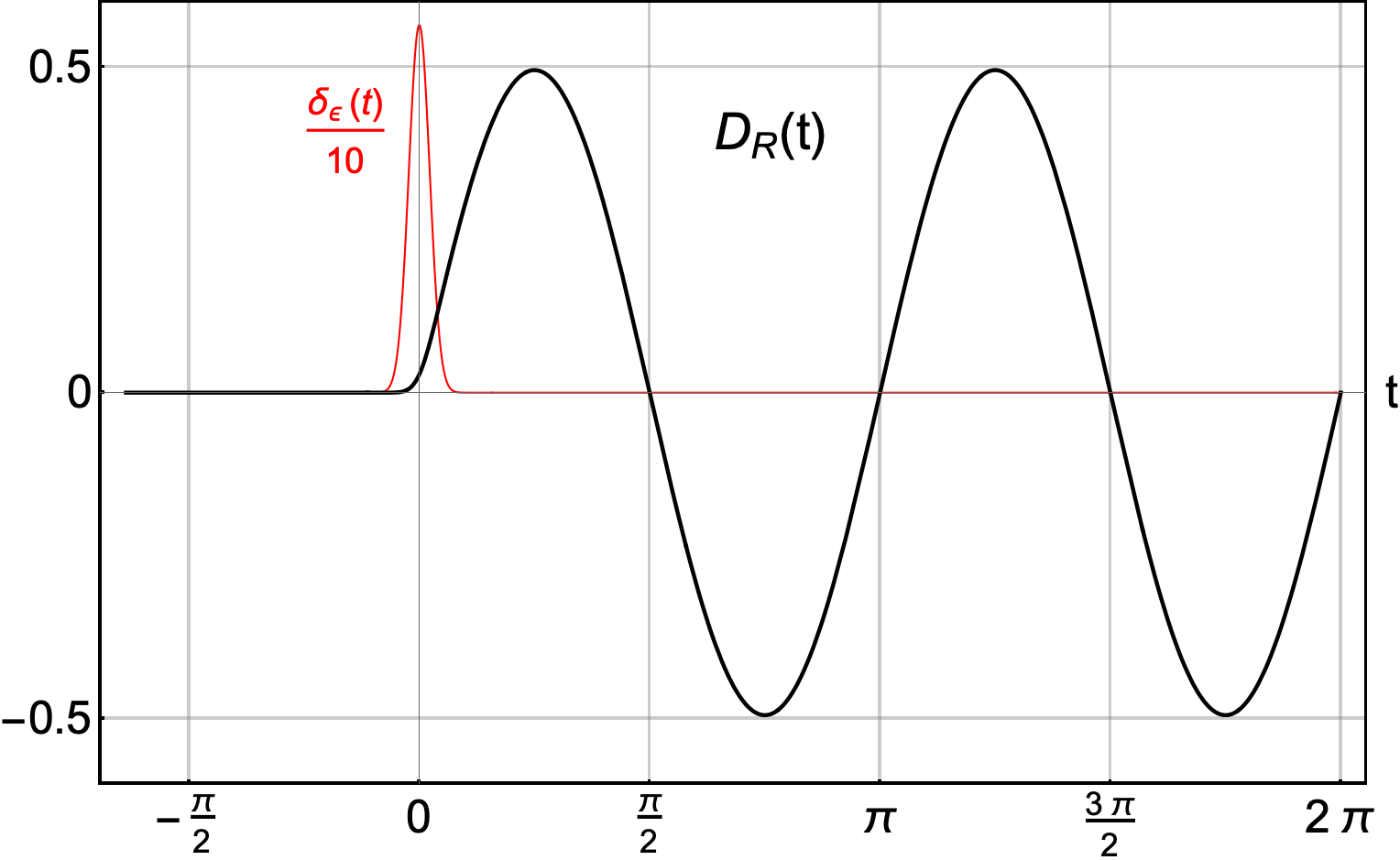}\label{fig:3a}}\hskip0.9cm
\subfigure[]{\includegraphics[width=0.45\textwidth,height=0.25\textwidth, angle =0]{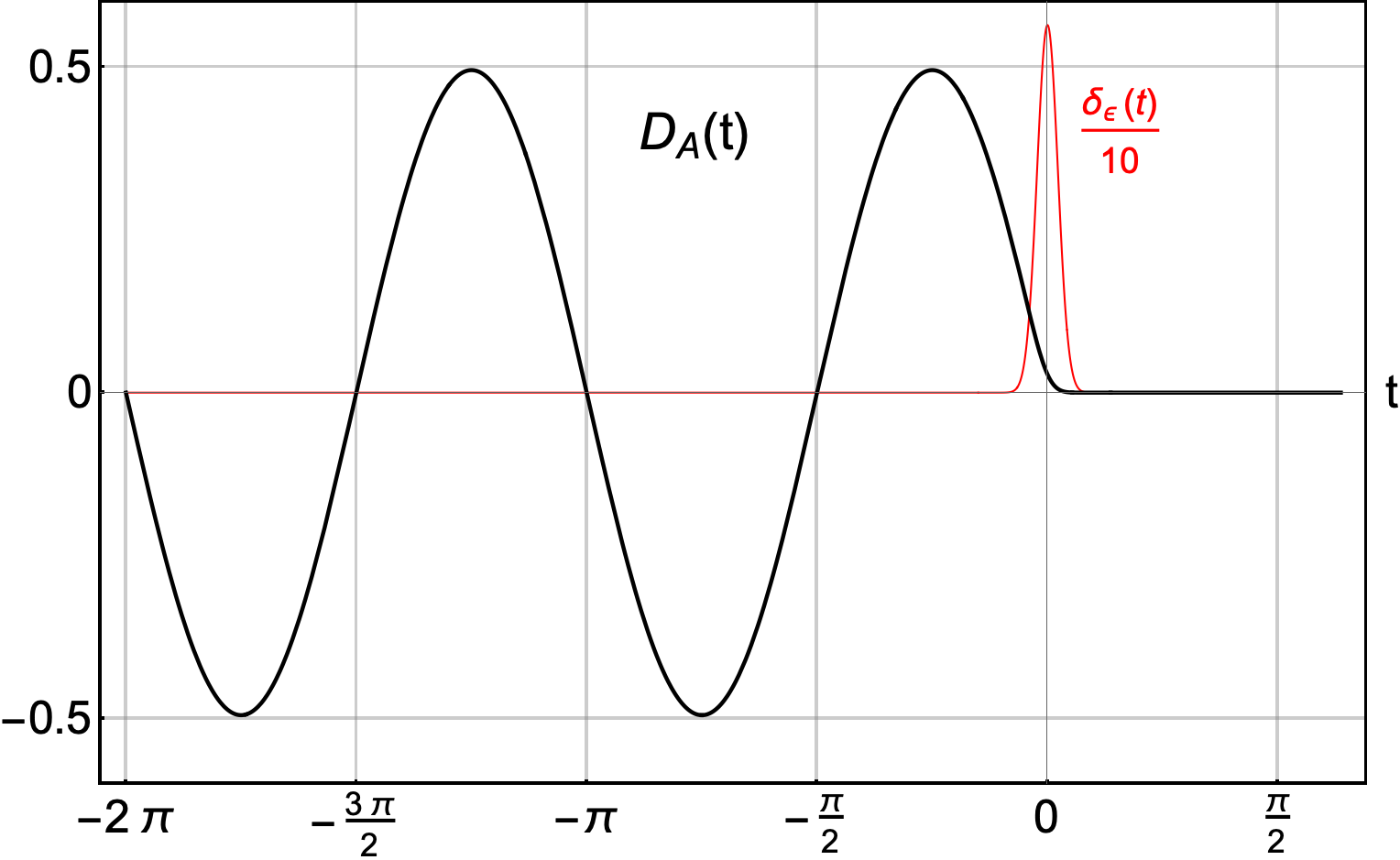}\label{fig:3b}}
\caption{Fundamental harmonic oscillator numerical solutions. (a) Retarded solution $D_R(t)$ and (b) advanced solution $D_A(t)$ for $\omega =2.0$, $\epsilon=0.005$. For better visibility, in both panels, the regularized delta force  $\delta_{\epsilon}(t)$ is divided by factor 10.}
\end{figure}

We will analyze a retarded solution initialized with specific conditions at a designated time point, $t_0$. Subsequently, we will solve the equation for times greater than or equal to $t_0$. We select initial conditions, $D(t_0)=0$ and $D'(t_0)=0$, which would yield a null solution in the absence of a delta force. The introduction of a delta force produces a non-trivial solution, illustrated in Figure \ref{fig:3a}. Figure \ref{fig:3b} follows a similar approach for the advanced solution $D_A(t)$. Our numerical results can be compared with well-known exact solutions $D_R(t)=\theta(t)\frac{\sin(\omega t)}{\omega}$ and $D_A(t)=-\theta(-t)\frac{\sin(\omega t)}{\omega}$. 
These results demonstrate that the system receives a burst of energy around $t = 0$, leading to oscillations. Outside this limited interval, the influence of the force becomes negligible, and the oscillations are governed by the same equation that applies in cases with non-zero initial conditions.

In contrast to the harmonic oscillator example, the situation with shock waves presents a critical distinction. In this case, the variable $z$ does not correspond to time. As a consequence, the delta force is not impulsive, acting at a single instant, but rather exhibits continuous presence at the light cone surface. This signifies a sustained influx of energy into the system.

The exact solution suggests using initial conditions $\phi(z_0)=0$ and $\phi'(z_0)=0$ at some $z_0$ greater than zero, followed by solving the equation for $z$ less than or equal to $z_0$. However, the numerical solution encounters difficulties due to a singularity at $z = 0$. This singularity results in unbounded growth of both the function $\phi(z)$ and its derivative.

As shown in equations \eqref{eq:sol-Wk-n=1} and \eqref{eq:sol-Wk-n=23}, the analytical solution for various dimensions involves singular terms ($\ln(-z)$, ${(-z)^{-1/2}}$, and $(-z)^{-1}$). These singularities are eliminated from $W_0(z)$ by setting a specific parameter ($\beta_0=0$). This ensures that the partial solutions around $z = 0$ become linear functions (meaning the second derivative is zero). Consequently, by restricting the numerical analysis to a small region around $z = 0$, we circumvent the problematic term involving the second derivative.  However, caution is necessary.

To address this, we split equation  \eqref{Fform} into two parts: $F(\phi)=F_1(\phi)+F_2(\phi)$, where $F_2(\phi)\equiv-z\phi''$ involves the troublesome second derivative term, while $F_1(\phi)\equiv-\frac{n+1}{2}\phi'+{\rm sgn}(\phi)$ doesn't. Since $F_2(\phi)$ acting on the solution $\theta(-z) W(z)$ results in a Dirac delta term
\[
F_2\big(\theta(-z) W(z)\big)=-W(0)\delta(z)-\theta(-z)zW''(z),
\] 
the coefficient multiplying the delta term on the right side of the equation \eqref{eq:L} also needs adjustment when $F_2$ is ommited.
Finally, evaluating $F_1(\phi)$ on the solution $\theta(-z) W(z)$ yields another specific result
\[
F_1\big(\theta(-z)W(z)\big)=\frac{n+1}{2}W(0)\delta(z)+\theta(-z)\Big[-\frac{n+1}{2}W'(z)+{\rm sgn}(W)\Big]. 
\]
Our analysis of $F(\phi)$ reveals that replacing it with $F_1(\phi)$ on the left side of the equation \eqref{eq:L} necessitates a corresponding change on the right side containing delta term. This adjustment involves replacing the coefficient $\frac{n-1}{2}$ with $\frac{n+1}{2}$ for consistency. As a consequence, we obtain a new equation near $z=0$:
\be
-\frac{n+1}{2}\phi'+{\rm sgn}(\phi)=\frac{n+1}{2}W(0)\delta(z).\label{reducedeq}
\ee
This approach demonstrates that even for $n=1$, we still end up with a non-homogeneous equation close to $z=0$. This implies the possibility of obtaining a non-trivial solution despite initial conditions: $\phi(z_0)=0$, $\phi'(z_0)=0$. This outcome validates our strategy.

We now proceed with the numerical solution of equation (previously referred to as the reduced equation) for the specific case of $n = 2$. We will treat $W(0)$ as a free parameter and use a regularized version (equation referred to as the regularized delta) to approximate the exact Dirac delta function.

Here is the breakdown of our approach:
We select a point, $z_0 = 0.3$, located outside the light cone. The initial conditions at this point are $\phi(z_0)=0$ and $\phi'(z_0)=0$. In the absence of the delta force term, the only solution to the equation with these initial conditions would be zero.
We introduce the regularized delta force with a specific parameter $\epsilon = 10^{-6}$. We also fix the free parameter $W(0)$ at a value of $2/3$.
Then, we solve the equation numerically on a symmetrical interval ranging from $-z_0$ to $z_0$ (i.e., $-0.3 \le z \le 0.3$).  

\begin{figure}[h!]
\subfigure[]{\includegraphics[width=0.45\textwidth,height=0.25\textwidth, angle =0]{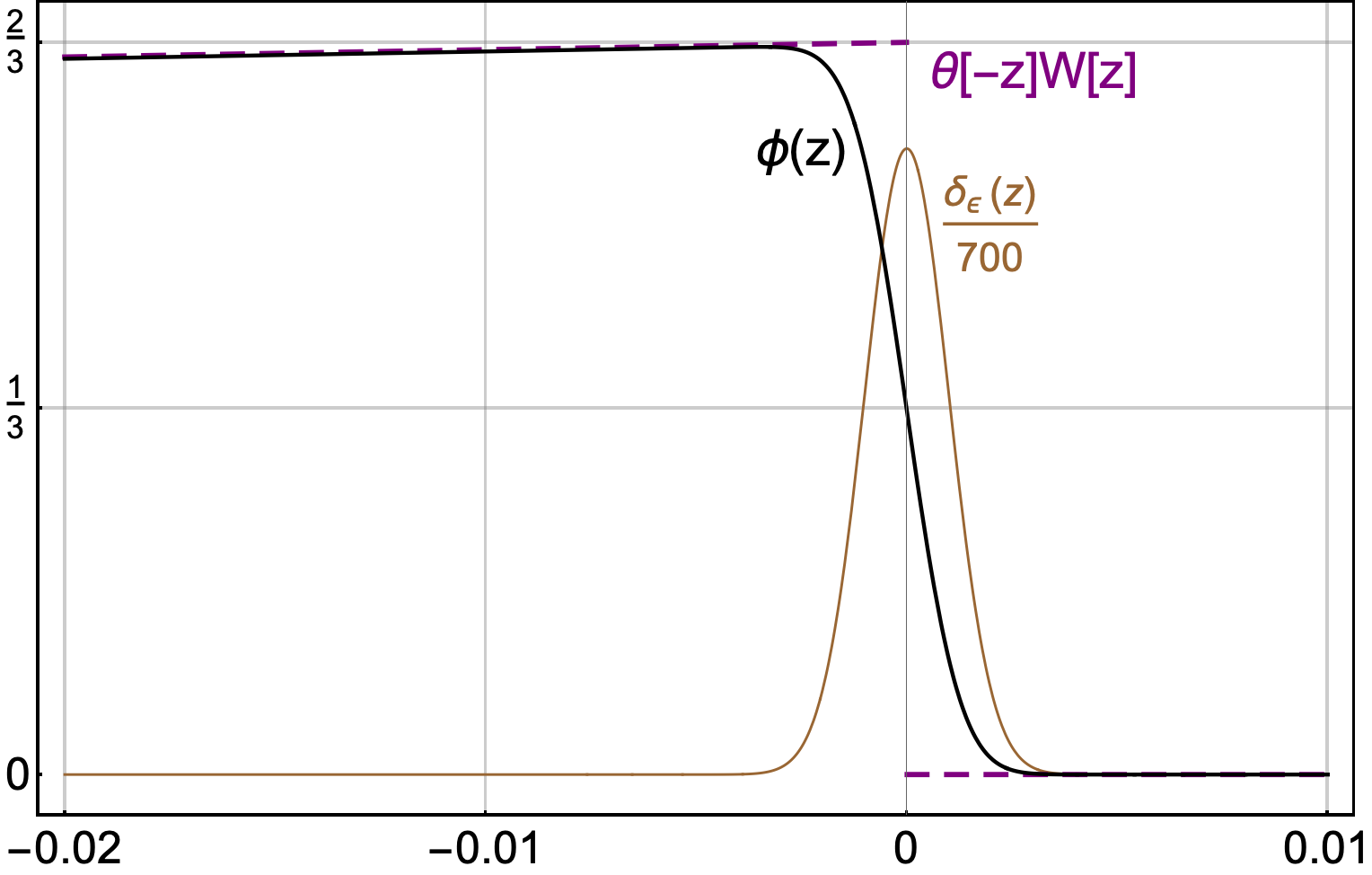}\label{fig:4a}}\hskip0.9cm
\subfigure[]{\includegraphics[width=0.45\textwidth,height=0.25\textwidth, angle =0]{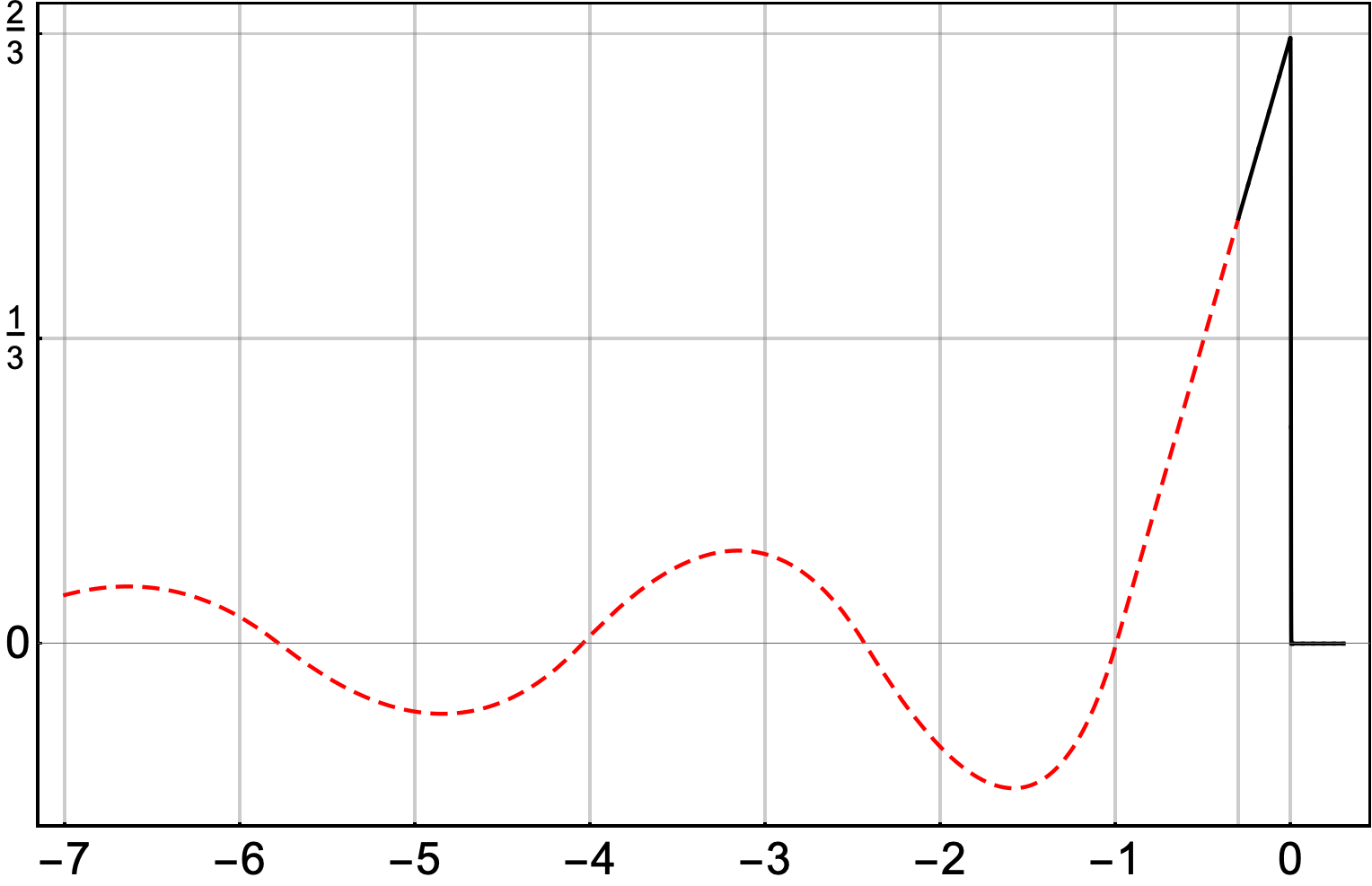}\label{fig:4b}}
\caption{The numerical solution for $\phi(z)$ in equation \eqref{eq:L} for the case $n=2$, $W(0)=\frac{2}{3}$ and the delta function approximated using a regularized version $\delta(z)\approx\delta_{\epsilon}(z)$ ($\epsilon=10^{-6}$).  (a) Enlarged view of the solution near $z=0$.  (b) Complete solution obtained by solving equation  \eqref{reducedeq}  first, followed by solving equation  \eqref{eq:L}. The matching point between the two solutions is $z_m=-0.3$.}
\end{figure}
Figure \ref{fig:4a} depicts only a tiny element of this solution. We have also displayed a delta force to indicate the region where it is relevant. To keep the plot's scale, the delta $\delta_{\epsilon}(z)$ was shown after being divided by some arbitrary factor. We also plot a piece of the $\theta(-z)W_0(z)$ solution to compare it to a numerical solution. In a tiny region about $z=0$, when delta force is significantly different from zero, the numerical solution exhibits nonlinear behavior. It fast approaches the analytical partial solution  $W_0=\frac{2}{3}(z+1)$, which is denoted by a dashed line. In Fig.\ref{fig:4b}, we additionally depict a part of a numerical curve at $z<-z_0$, which requires solving the whole equation \eqref{eq:L}. 

To obtain the solution for $z$ less than $z_0$, we utilize the values of $\phi(-z_0)$ and $\phi'(-z_0)$ derived from solving equation \eqref{reducedeq} as our initial data. The red dashed curve in Figure \ref{fig:1a} represents this second part of the solution.

As it can be seen, the numerical solution closely matches the actual behavior depicted in Figure \ref{fig:1a}. Furthermore, numerical analysis demonstrates that adjusting the free parameter $W(0)$ that multiplies the delta force allows us to achieve the desired maximum (discontinuity in the exact solution) at the light cone.


\section{Numerical Investigation of Shock Waves in (2+1) Dimensions} \label{sec:numeric}

\subsection{General remarks}

This section outlines the technical aspects of the numerical simulation. Standard explicit Runge-Kutta method with fourth-order time oversampling (RK4) is used for integrations. The field state is represented by numerical values (double-precision) distributed across a rectangular mesh.
The simulation focuses on a square region with side length $L$, chosen as a subset of the simultaneity surface centered on the shock wave in its reference frame. This square is further divided into a grid of $N \times N$ cells, resulting in a total of $N^2$ mesh points. Each cell has an area of $h^2$ (where $h = L/N$).
The entire field distribution is advanced through an RK4 step to calculate incremental changes in the field state from time $t$ to time $t + \delta t$. A fixed timestep ($\delta t = h/10$) is used for the simulation.

The mathematical operation involving second-order spatial derivatives (Laplace operator) is converted into a matrix representation suitable for computer calculations. This is achieved by applying a convolution process across the spatial mesh using a predefined kernel
\be 
[\Delta^2]_{kern} = \dfrac1{h^2}
\begin{bmatrix}
0 \quad  &  1 \quad  &  0 \\
1 \quad  & -4 \quad  &  1 \\
0 \quad  &  1 \quad  &  0
\end{bmatrix}.
\ee

It's important to acknowledge that the chosen square-shaped mesh with uniform spacing might not be ideal for radially symmetric solutions like shock waves. This can lead to a discrepancy between the theoretical (continuous) isotropic nature of the boundary conditions and the actual anisotropic (and potentially discontinuous) behavior captured by the numerical simulation, especially at lower grid resolutions ($N$).
In theory, as the grid size gets infinitely large ($N\rightarrow \infty$), the numerical representation becomes more isotropic, but for practical simulations, this may not be achievable. This mismatch can introduce errors that affect the radial symmetry of the solution. The observed deviation from radial symmetry serves as an indicator of the accumulated numerical error and is used to assess the validity of the simulation.



\subsection{Shock wave solution for the Dirichlet boundary condition}

As demonstrated in sections \eqref{sec:energy}, \eqref{sec:en2+1} and \eqref{sec:en3+1}, the exact solutions for these waves do not conserve energy. In simpler terms, the energy within the light cone (where the solution deviates from zero) increases over time (refer to equation \eqref{func:general_sw_energy}for details).
This continuous energy influx within the light cone is necessary for the exact solution to exist. 

In this section, we aim to explore mimicking the impact of the delta force by explicitly implementing boundary conditions. 
Given that the exact solution for a shock wave coincides with a constant value at the light cone, we can attempt to reproduce this behavior by solving the homogeneous SG equation and explicitly imposing a Dirichlet boundary condition at the light cone. The region of interest for solving the equation is an $n$-dimensional ball with a radius of $r = t$. The Dirichlet condition serves to transfer energy into the ball where the shock wave resides.

To illustrate this concept further, consider a two-dimensional ($n = 2$) membrane in a gravitational field. Initially, the membrane is raised to a height of W(0) above a rigid floor and extends infinitely. An external force maintains this initial position. In this state, the membrane possesses infinite gravitational energy but lacks any gradient or kinetic energy.

Now, imagine releasing the force holding the membrane within a circular region that starts at a point $r = 0$ and expands with a radius growing as $r = t$. During this process, the energy contained within the circle becomes equal to the energy density $V(\phi) = |\phi| = W(0)$ (the value of the field at the light cone) multiplied by the area of the circle. This precisely reflects the energy of a shock wave in (2 + 1) dimensions.

Within the expanding circular region, the field is free to evolve, and its behavior is governed by the homogeneous SG equation. While this analogy is particularly intuitive for $n = 2$ dimensions, the same concept can be applied to higher spatial dimensions. This approach is particularly advantageous for numerical solutions as it circumvents the need to implement a regularized (and consequently, approximated) delta force term.

\begin{figure}[h!]
		\centering
		\includegraphics[width=0.6\textwidth,height=0.3\textwidth, angle =0]{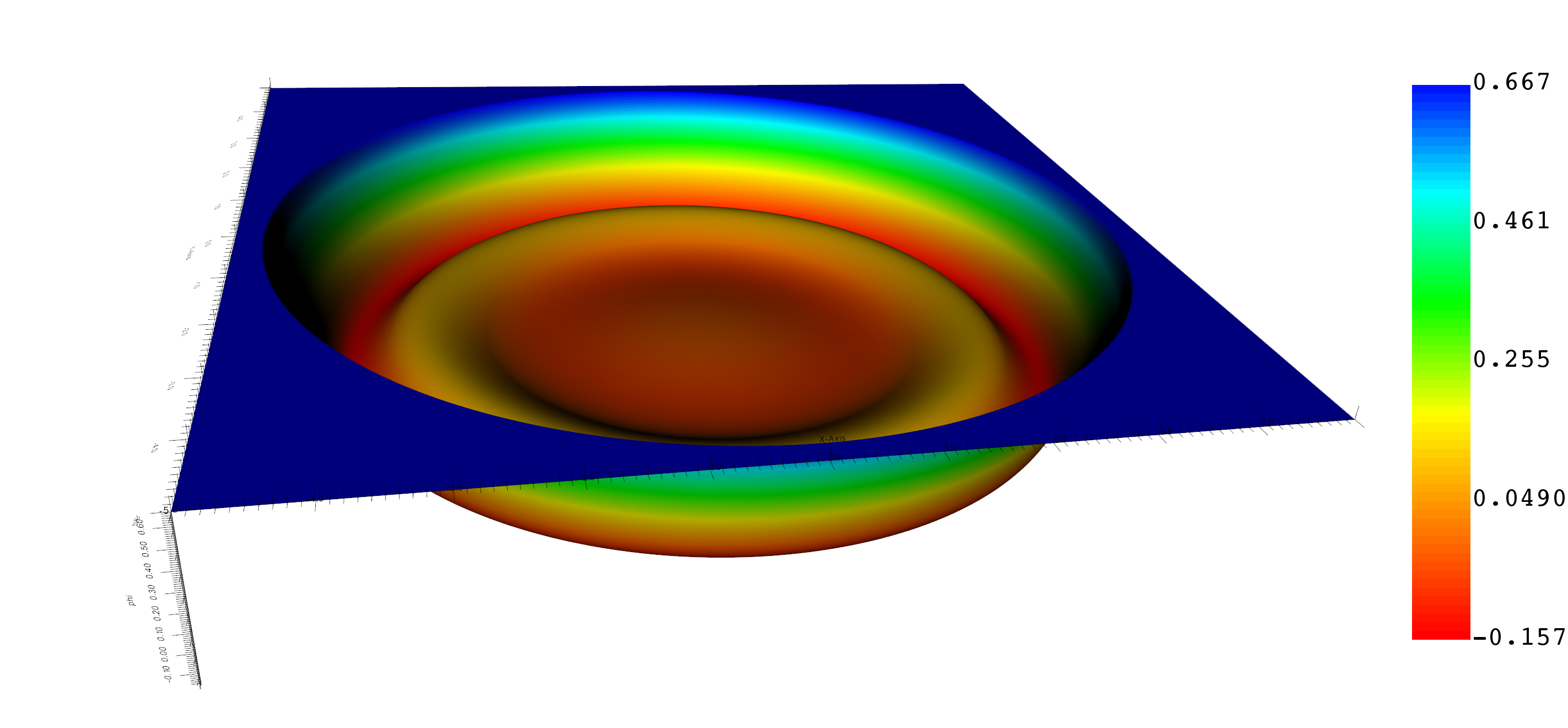}
		\caption{Numerical evolution in a scenario where the field value is initially constant at $W(0)=\frac{2}{3}$ (represented by the plateau) throughout the entire space (except for the light cone region). Within the light cone of the event $(t, r) = (0, 0)$, the field is released to allow for its dynamics. The colorbar and height represent the field value. The underlying numerical mesh discretization has a different symmetry compared to the ideal analytical solution. However, the lack of significant impact on the simulation suggests minimal accumulation of numerical error.}
		\label{fig:field_up}
\end{figure}

An illustrative example of a numerical simulation for the $n = 2$ case that employs this concept is presented in Fig.~\ref{fig:field_up}. This solution is in complete agreement with our exact solution.

In technical terms, the specific value of the field outside the light cone is inconsequential for the solution's dynamics, which depend solely on the field's value at the light cone. Consequently, it is permissible to consider a scenario where the field is identically zero outside the light cone while retaining its value of $W(0)$ at the light cone. The application of a Dirichlet boundary condition to the homogeneous equation within the light cone produces an effect on the field that is identical to the influence of a delta function.

We solve equation \eqref{eqsg1} within the region $r < t$, imposing a constant value boundary condition at the light cone surface $r=t$. This specific type of boundary condition defines the class of solutions we are investigating, which is represented by the initial guess (ansatz) given in equation \eqref{eq:req-discont}.
The simulation is confined to the region $r < t$.  The behavior of the surface at the light cone ($r = t$) and the field outside this boundary ($r > t$) are not explicitly calculated in the simulation but are considered to be fixed with specific, predetermined values
\begin{eqnarray}\label{Dirichlet}
\phi(t,\vec x)=\left\{
\begin{array}{ccc}
W(0)&{\rm for}&r=t\\
0&{\rm for}&r>t
\end{array}
\right..
\end{eqnarray}

\begin{figure}[h!]
	\centering
	\includegraphics[width=0.6\textwidth,height=0.3\textwidth, angle =0]{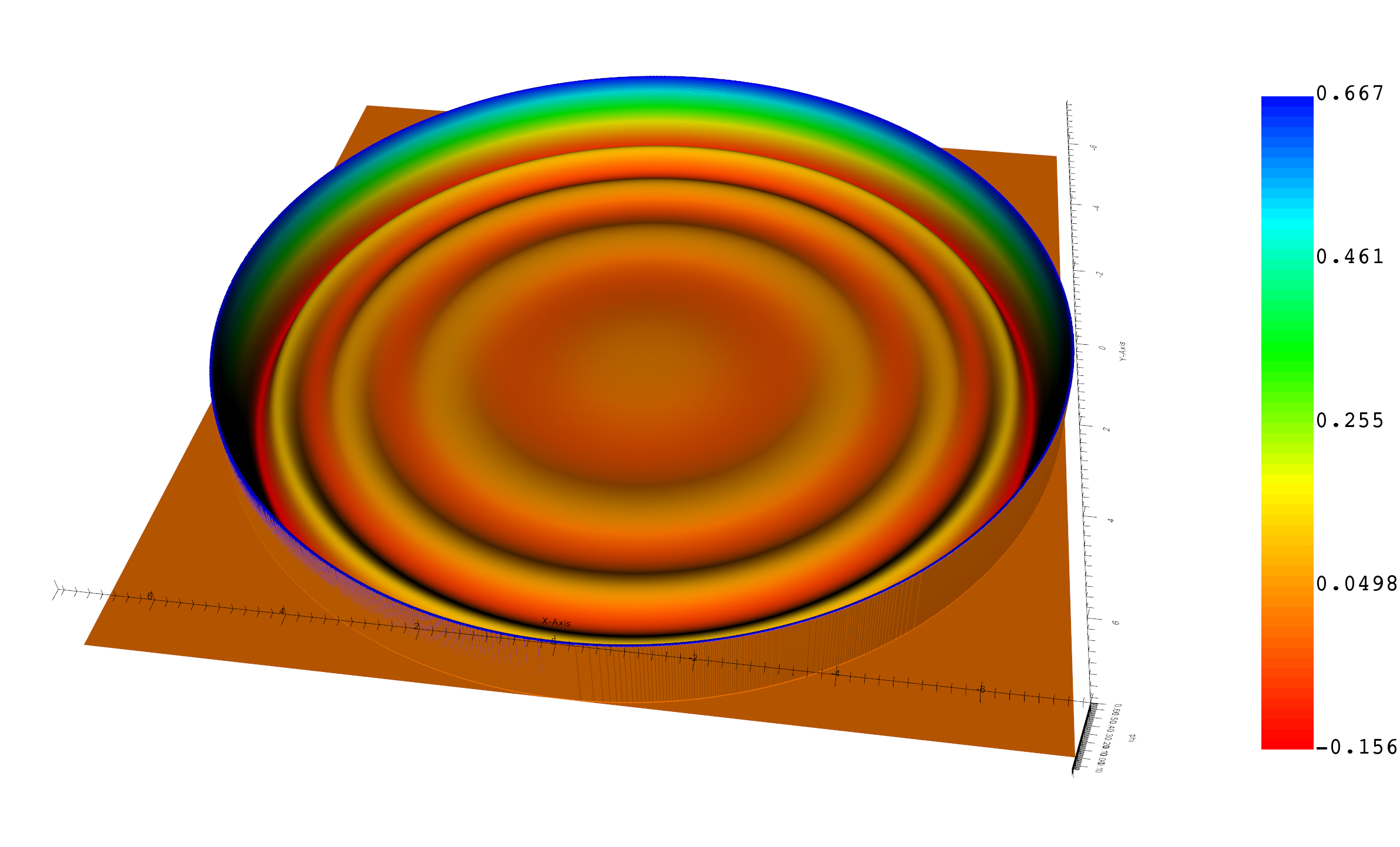}
	\caption{The reference simulation for a shock wave with a discontinuity of $W(0) = 2/3$ at a time $t = 7.5$. The colorbar and height represent the field value.}
	\label{fig:fullspace_shockwave}
\end{figure}
\begin{figure}[h!]
	\centering
	\subfigure[]{\includegraphics[width=0.48\textwidth,height=0.24\textwidth, angle =0]{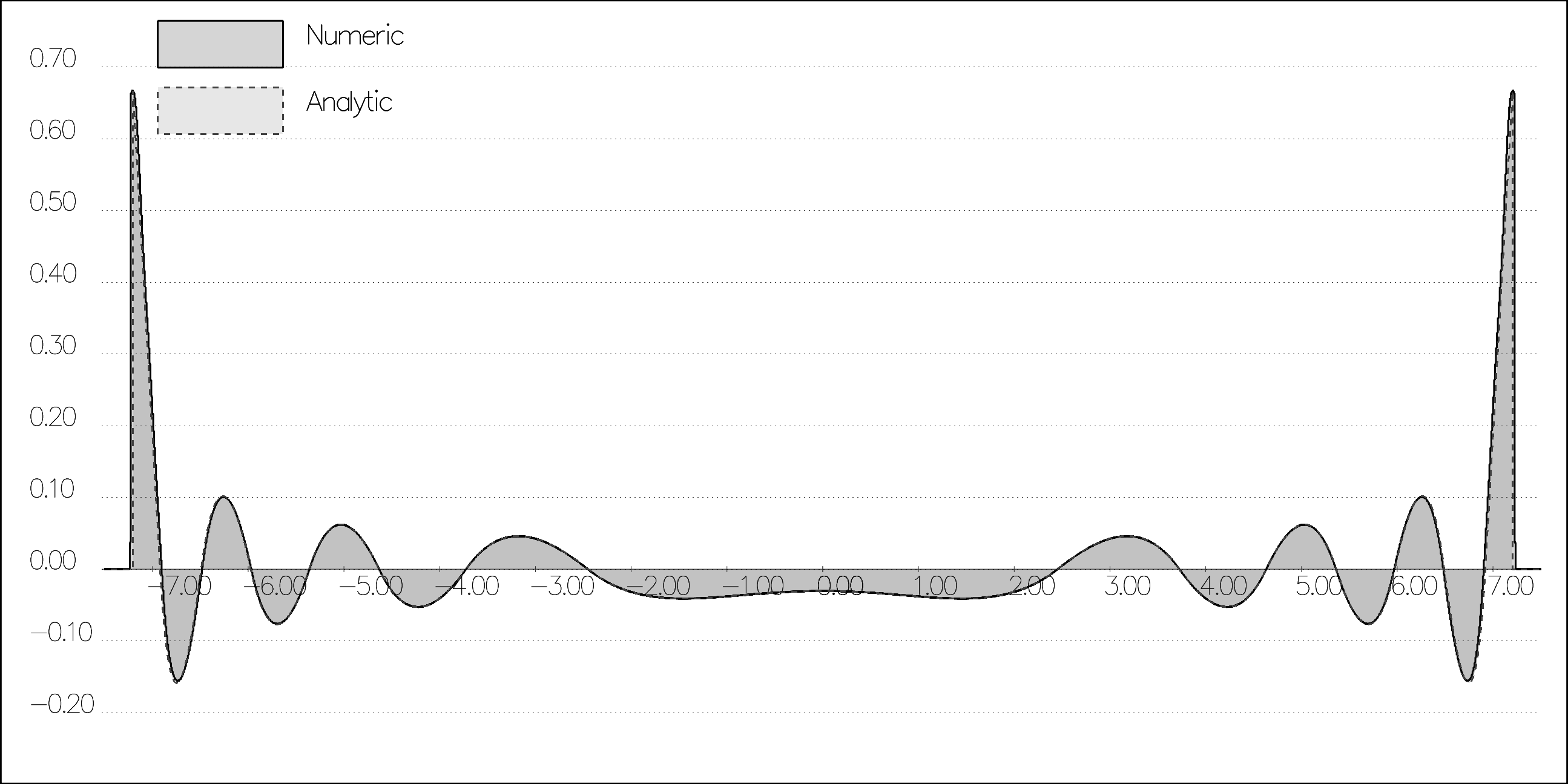}\label{fig:dirichlet1a}}
	\subfigure[]{\includegraphics[width=0.48\textwidth,height=0.24\textwidth, angle =0]{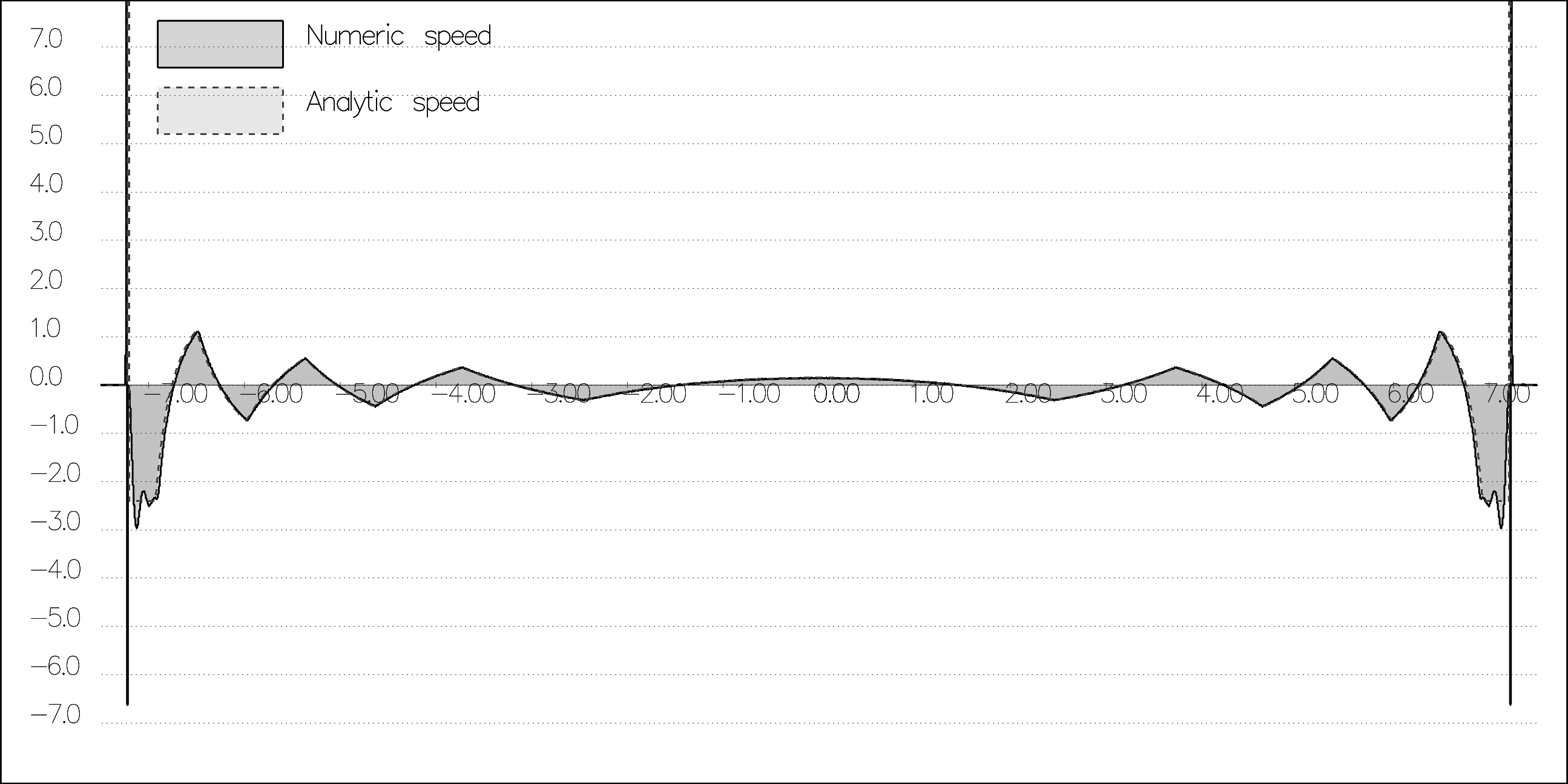}\label{fig:dirichlet1b}}
	\caption{(a) A comparison between a cross-section of the shock wave profile obtained from our simulation (at $y = 0$, $t = 7.25$) and the analytical solution for the shock wave. (b) The time derivative of the shock wave profile from both the numerical solution (at $y = 0$, $t = 7.25$) and the analytical solution.}
\end{figure}

The results of a numerical simulation for the SG field are presented in Figure~\ref{fig:fullspace_shockwave}. A detailed analysis of the shock wave at time $t = 7.25$ is provided in Figure~\ref{fig:dirichlet1a} Panel (a) shows a cross-section of the wave profile along the $x$-axis, extracted from the simulation data in Figure \ref{fig:fullspace_shockwave}. Panel (b) depicts the corresponding time derivative of the wave profile.

\begin{figure}[h!]
	\centering
	\subfigure[]{\includegraphics[width=0.49\textwidth,height=0.245\textwidth, angle =0]{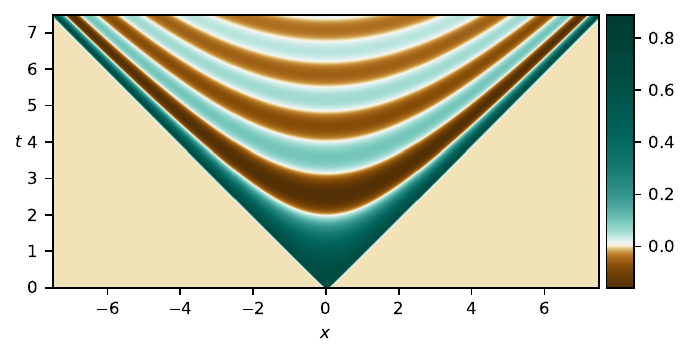}\label{fig:dirichlet2a}}
	\subfigure[]{\includegraphics[width=0.49\textwidth,height=0.245\textwidth, angle =0]{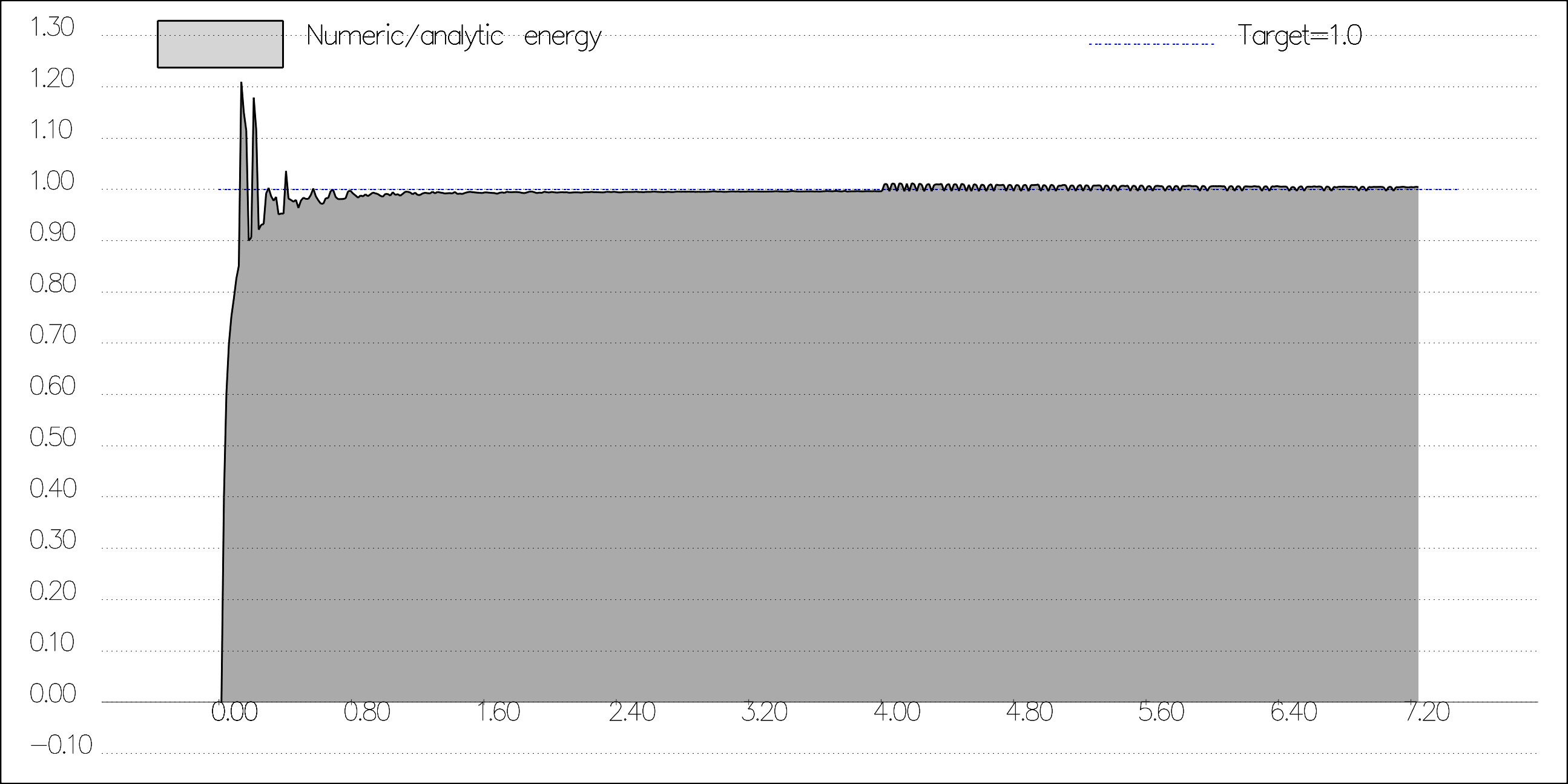}\label{fig:dirichlet2b}}
	\caption{The behavior of the shock wave solution in two spatial dimensions ($n = 2$). Panel (a) shows a color-coded spacetime section ($tx$-plane) representing the gradient of the shock wave solution. Panel (b) depicts the time evolution of the ratio between the energy obtained from the numerical simulation and the theoretical energy (analytical solution) within the shock wave region. The initial spikes in this ratio are likely due to numerical errors associated with including or excluding specific grid points (discrete square sites) during the energy integration process in the simulation.}
	\label{fig:dirichlet2}
\end{figure}

Figure \ref{fig:dirichlet2a} presents a spacetime diagram depicting the evolution of the shock wave solution. The color gradient indicates the value of the scalar field. The plane shown is the $tx$-plane (time vs. $x$-axis). Fig.\ref{fig:dirichlet2b} shows the time evolution of the ratio between the numerical solution obtained in our simulation and the analytical solution for the shock wave.

\subsection{Shock wave disintegration}

This section investigates the behavior of a shock wave after its energy source is turned off.  We consider a shock wave generated by the evolution of a scalar field $\phi$ within a specific time interval ($0<t<t_0$) under the constraint \eqref{Dirichlet}. At time $t=t_0$, this constraint is disabled, analogous to removing a delta force term in a non-homogeneous equation. Disabling this condition disrupts the energy transfer process to the shock wave, preventing it from propagating according to analytical predictions.

\begin{figure}[h!]
	\centering
	\includegraphics[width=0.75\textwidth]{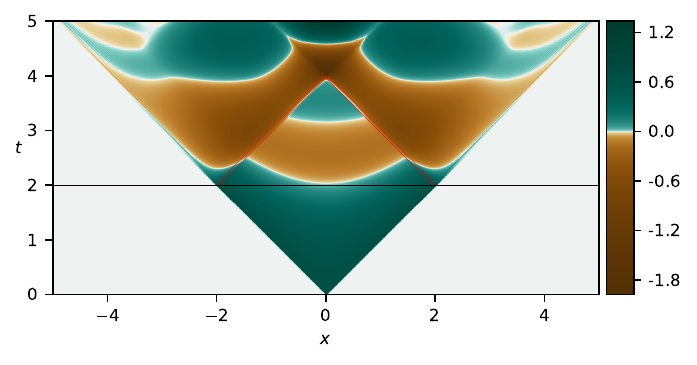}
	\caption{The initial stages of the shock wave weakening after its energy source is disabled. The horizontal black line indicates the moment ($t = 2.00$) when condition \eqref{Dirichlet}, analogous to a delta force term, is turned off.
The red diagonal lines represent the boundary between the region influenced by the delta force (before $t = 2.00$) and the region where it's no longer active (after $t = 2.00$). These lines appear in closely spaced, parallel pairs. The separation between each pair is $2\epsilon$, where $\epsilon$ is a parameter related to the specific method used to represent the delta function.}
	\label{fig:dirichlet3closeup}
\end{figure}

\begin{figure}[h!]
	\centering
	\subfigure[\hskip 0.1 cm Section $tx$ of $(2+1)$ dimensional Minkowski diagram]{\includegraphics[width=0.48\textwidth]{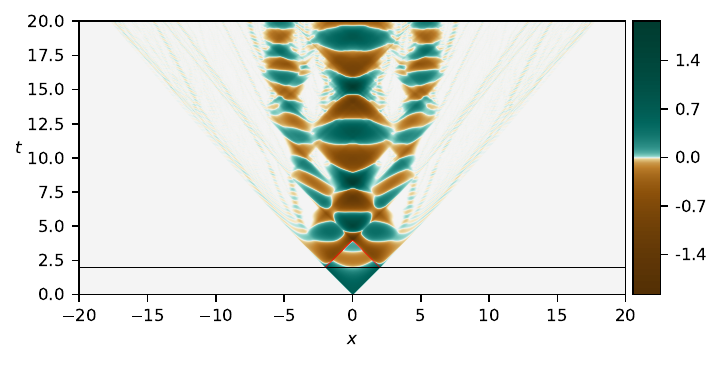} \label{fig:dirichlet3a}}
	\subfigure[\hskip 0.1 cm  Section with $22.5\deg$ ($\pi/8$) rotation from that of (a)]{\includegraphics[width=0.48\textwidth]{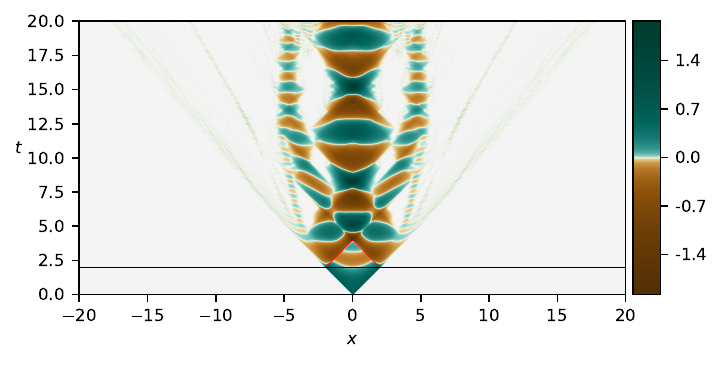} \label{fig:dirichlet3b}}
	\subfigure[\hskip 0.1 cm Intermediate snapshot of the center of the field at $t=10$, with side $L=10$]{\includegraphics[width=0.48\textwidth]{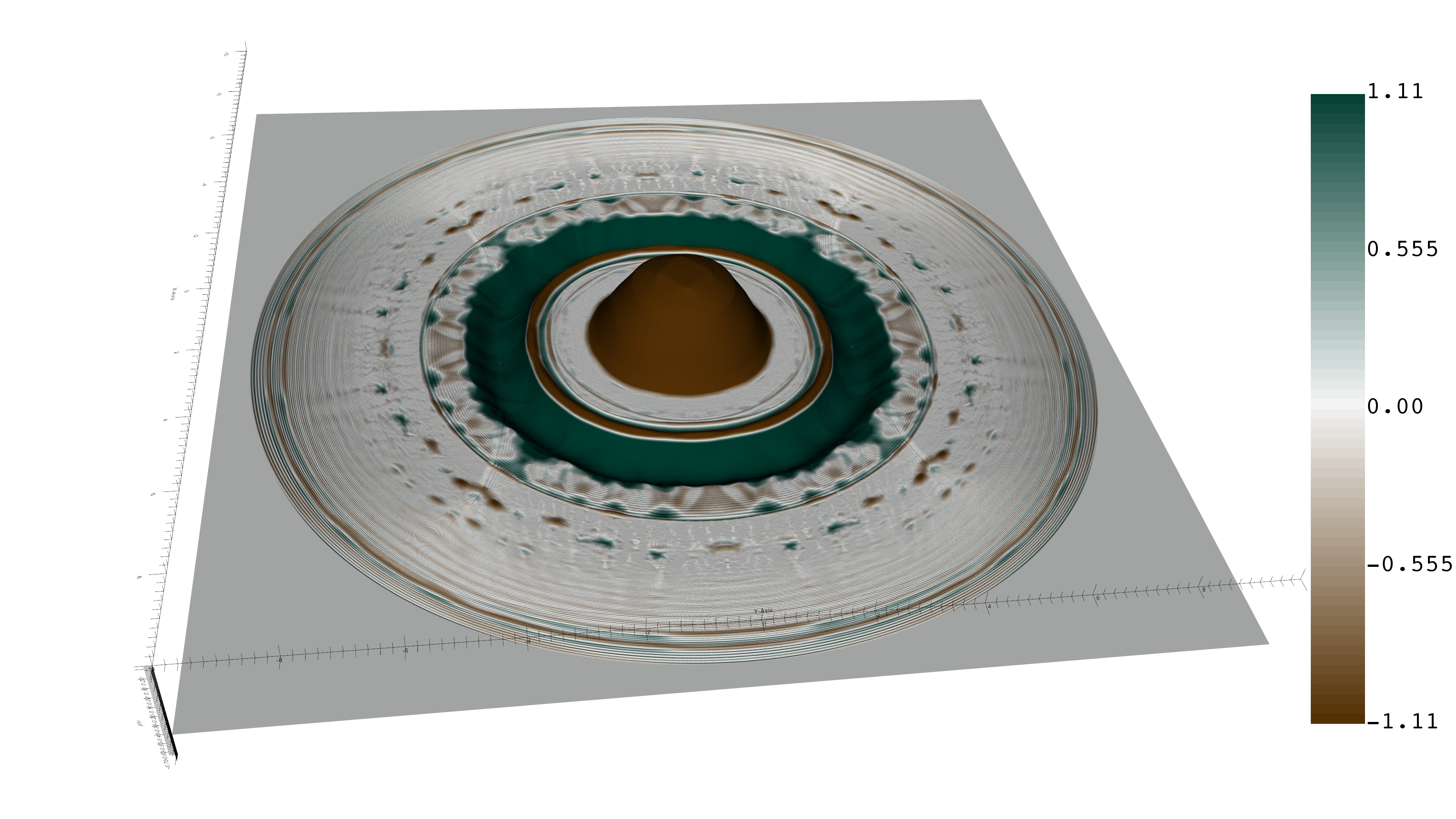} \label{fig:dirichlet3c}}
	\subfigure[\hskip 0.1 cm The final full field configuration, at $t=20$, showing all simulation region with side $L=20$]{\includegraphics[width=0.48\textwidth]{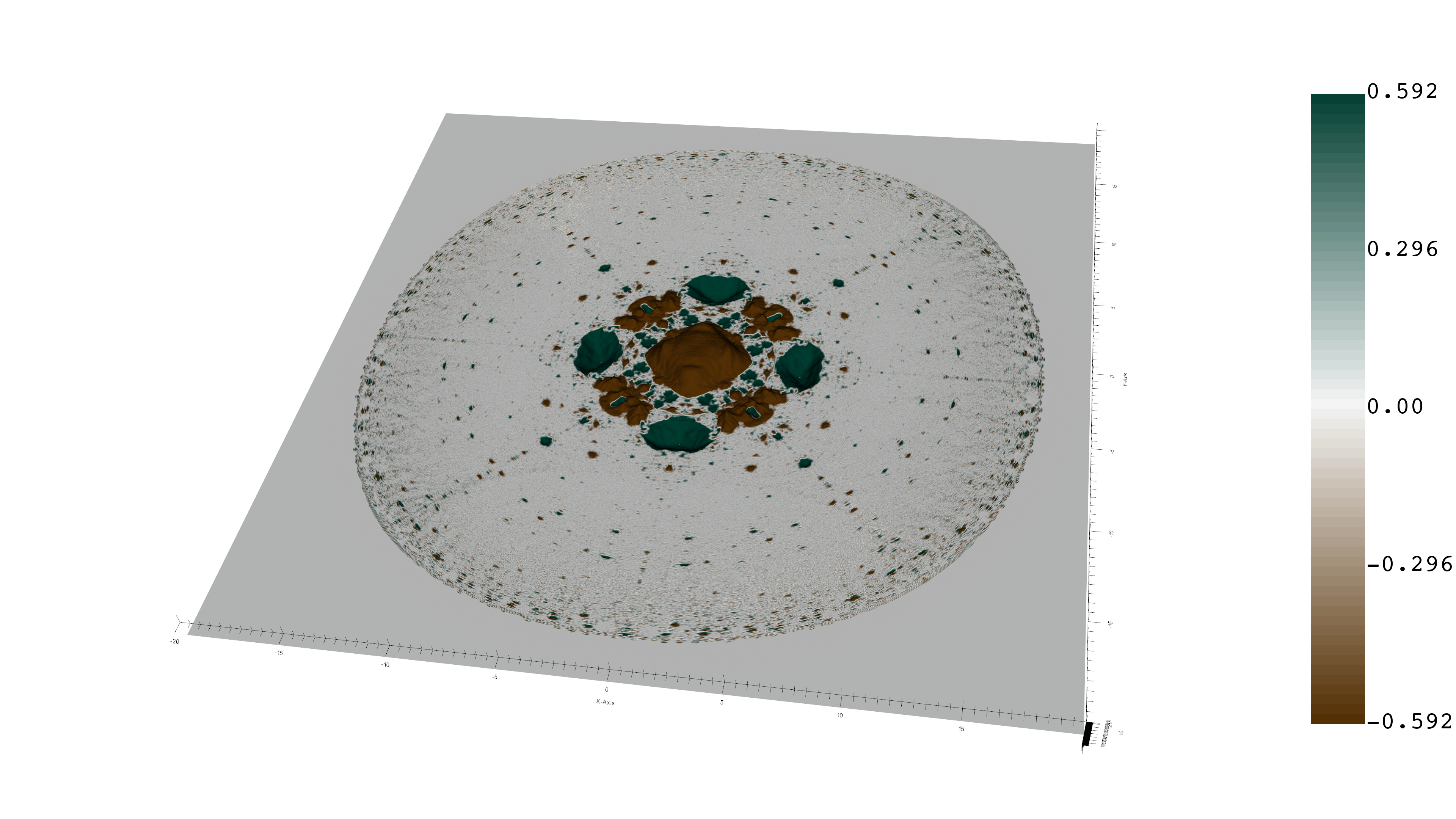} \label{fig:dirichlet3d}}
	\caption{ Panels (a) and (b) depict the initial stages of the shock wave weakening after its energy source is disabled. The horizontal black line indicates the moment ($t = 2.00$) when condition \eqref{Dirichlet}, analogous to a delta force term, is turned off.
The red diagonal lines represent the boundary between the region influenced by the delta force (before $t = 2.00$) and the region where it's no longer active (after $t = 2.00$). These lines appear in closely spaced, parallel pairs. The separation between each pair is $2\epsilon$, where $\epsilon$ is a parameter related to the specific method used to represent the delta function.
Panels (c) and (d) show snapshots of the field at later times: $t = 10$ (c) and $t = 20$ (d). Notice the gradual loss of symmetry (isotropy) in the field distribution as time progresses, particularly evident in panel (d).}
	\label{fig:dirichlet3}
\end{figure}

Figure \ref{fig:dirichlet3closeup} provides a close-up view of the field's history in the $tx$-plane. This figure zooms into the lower central region of Figure \ref{fig:dirichlet3a}.  It shows that within the intersection of the future light cone (originating from event $(0, 0)$) and the past light cone (originating from event $(2t_0,0)$), the shock wave remains unaffected.

Figure \ref{fig:dirichlet3} showcases this phenomenon in various planes of (2+1) Minkowski space.
Figure \ref{fig:dirichlet3a} shows the history of the field at $y = 0$, essentially how this plane of the field evolves over time. Figure \ref{fig:dirichlet3b} depicts a view at an angle of $\pi/8$ from Figure \ref{fig:dirichlet3a}, corresponding to the $y = x/2$ plane whereas Figure \ref{fig:dirichlet3c} zooms in on the central part of the $t = 10$ plane, representing a snapshot of the field's configuration at the simulation's midpoint ($t = 10$). Finally, Figure \ref{fig:dirichlet3d} shows the final field configuration at the end of the simulation ($t = 20$).

Similar to observations in one spatial dimension ($n = 1$) \cite{Hahne:2019odw}, our simulations in two dimensions ($n = 2$) reveal the formation of compact, localized structures within the field. These structures include a central, bump-shaped feature resembling the analytical oscillon and a novel ring formation surrounding the central bump. While Figures \ref{fig:dirichlet3a} and \ref{fig:dirichlet3b} suggest some temporal consistency in these structures, Figures \ref{fig:dirichlet3c} and \ref{fig:dirichlet3d} provide a clearer view of their shape. These later figures indicate that the central oscillating structure exhibits stability over time, while the compact ring breaks down into multiple, smaller oscillating structures.

The observed breakdown of rotational symmetry in the field is likely due to numerical errors that introduce high-frequency oscillations. However, we believe this doesn't negate the key findings: the instability of the ring structure and the stability of the central oscillating feature (oscillon). In fact, these errors might act as a strong perturbation, further revealing important properties of these structures.
The main large-scale structures in \ref{fig:dirichlet3c} at ($t=10$) -- the central bump resembling an oscillon and the surrounding ring -  maintain their rotational symmetry despite some added noise. This suggests that these structures play a significant role in the decay and thermalization of the shock wave in this two-dimensional ($2+1$) simulation.

\section{Conclusions and remarks}\label{sec:comments}

This study investigated the behavior of a specific class of solutions to the scalar field (SG) equation featuring a discontinuity near the light cone.  Previously, such solutions were only known to exist in one spatial dimension ($n = 1$).

Our work demonstrates that using the same ansatz in higher spatial dimensions leads to an ordinary equation with exact solutions. We successfully solved this equation for the cases of $n = 2$ and $n = 3$ spatial dimensions.  A surprising finding is that determining the solution zeros does not require any numerical methods for solving algebraic equations.  In the particular case of spherical $n = 3$ shock waves, finding these zeros is particularly straightforward.

Another key distinction is that shock waves represented by $\phi(z)=\theta(-z)W(z)$ are non-homogeneous solutions of the SG equation, incorporating an additional Dirac delta function localized at the light cone.  Fundamental solutions of differential equations fall into this category.

Beyond these core findings, we explored the behavior of shock waves after their energy source is disabled. Disabling the source (analogous to removing a delta force term) disrupts the energy transfer to the shock wave, preventing its propagation as predicted by analytical calculations. We identified the unaffected region within the light cones (Figure \ref{fig:dirichlet3closeup}).

The simulations in two dimensions ($n = 2$) revealed the formation of interesting structures within the field upon energy source removal.  These include a central, bump-shaped feature resembling an analytical oscillon and a novel ring formation surrounding it (refer to Figures \ref{fig:dirichlet3a} and \ref{fig:dirichlet3b} for initial observations). While the central oscillating structure exhibits stability over time, the compact ring breaks down into multiple, smaller oscillating structures. This breakdown might be partially attributed to numerical errors causing high-frequency oscillations in the field. However, the presence of these errors might also act as a strong perturbation, revealing important properties of the structures. The main large-scale structures observed at the simulation's midpoint ($t = 10$) maintain their rotational symmetry despite some noise, suggesting their role in the decay and thermalization of the shock wave (Figures \ref{fig:dirichlet3c} and \ref{fig:dirichlet3d}).

Overall, this study provides valuable insights into the behavior of shock waves under changing energy source conditions, the emergence of novel structures during the decay process, and the potential of the chosen ansatz for solving the SG equation in higher dimensions. While numerical errors might introduce some artifacts, they might also provide valuable information through perturbations. Future investigations could explore methods to mitigate these errors for even more precise results.

We conclude our discussion by drawing attention to a few key points:
\begin{enumerate}
\item
{\it Application to Forced SG Equations:}  For $n > 1$ (higher spatial dimensions), our solutions with $W(0) = 2$ might be useful in solving forced SG equations of the form $F(\phi(z))=f(z)$. Here, $F(\phi)$ represents the left-hand side of the SG equation in variable $z$, as given by equation \eqref{Fform}. Since the fundamental solution $D(z)=\theta(-z)W(z)$ satisfies $F(D(z))=\delta(z)$, the solution $\phi(z)$ for an arbitrary force $f(z)$ can be expressed as an integral involving $D(z)$ and $f(z)$.
\item
{\it Experimental Realizations:} While beyond the scope of this study, we emphasize the potential of investigating experimental realizations of shock waves in two or more spatial dimensions. The dynamics of the SG field in $n = 2$ within a gravitational field can be related to the motion of a two-dimensional elastic membrane over a rigid plane. The "unfolding transformation" \cite{Arodz:2005gz} accounts for reflection from the plane. The SG field acts as an auxiliary field, its dynamics mapped to membrane dynamics via an inverse folding transformation. By raising the membrane's altitude above the plane at $r = t$, we can replicate the effect of a delta force on the light cone. Furthermore, more general forces $f(z)$ could be considered to model raising the membrane's altitude and injecting energy into the system.
\item
{\it Shock Wave Disintegration and Structure Formation:} The existence of a shock wave solution requires uninterrupted energy flow into the region occupied by the wave. When this energy transfer ceases, the wave begins to disintegrate. In previous work, we observed that this degeneration in one spatial dimension ($n = 1$) resulted in the creation of numerous oscillon-like structures. Here, we pose two key questions in this context:
\begin{itemize}
\item {\it Question 1:} Are there dynamic processes that could generate structures similar to shock waves in 2+1 or 3+1 dimensions? This would be analogous to the process described in oscillon scattering \cite{Hahne:2019ela}, where we observed the development of shock waves in one spatial dimension (meaning an energy-conserving formation of very high-density energy reservoirs \textit{at} the solution's lightcone).
\item {\it Question 2:} How do shock waves break down in higher dimensions? Does this process produce semi-stable oscillons in higher dimensions? Unlike the one-dimensional case, there's currently no specific formula describing such oscillons. Our numerical study in this paper suggests that oscillon-like structures in 2+1 dimensions are more likely to resemble droplets than rings. Additionally, 2+1 dimensional shock waves degrade simultaneously in both radial and azimuthal directions. We believe a more comprehensive numerical investigation could provide valuable insights into this question.
\end{itemize}

\end{enumerate}
Overall, this study sheds light on the behavior of shock waves under changing energy source conditions, the emergence of novel structures during the decay process, and the potential of the chosen ansatz for solving the SG equation in higher dimensions. While numerical errors might introduce some artifacts, they might also provide valuable information through perturbations. Future investigations could explore methods to mitigate these errors for even more precise results. We also identified promising avenues for further research, including exploring the applicability of our solutions to forced SG equations and investigating experimental realizations of shock waves in higher dimensions.

\section*{Acknowledgmets}
The authors would like to thank the Open Source community for the tools used through this paper, in particular the Julia programming language community \cite{bezanson2012julia} and the C++ MFEM (finite elements) and VTK (visualization) libraries \cite{mfem,mfem-web,vtkBook}.
JSS would like to thank the support by CAPES Scholarship during part of the production of this article.


\bibliography{bibliography.bib}

\end{document}